\documentclass[12pt]{article}
\pdfoutput=1    
\usepackage{arxiv}
\usepackage{hyperref}
\usepackage{amsfonts}
\usepackage{amsmath}
\usepackage{amssymb}
\usepackage{amsthm}
\usepackage{graphics}
\usepackage{multicol}
\usepackage{multirow}
\usepackage{caption}
\usepackage{verbatim}
\usepackage{rotating}
\usepackage{array}
\usepackage{xcolor}
\usepackage{mathtools, cuted}
\usepackage{lscape}     
\usepackage{booktabs}     
\usepackage[header,title,page]{appendix}
\newtheorem{remark}{Remark}

\title{ASTERIX: Module for modelling the water flow on vegetated hillslopes}
\date{\today}

\usepackage{authblk}

\author[1]{Stelian Ion\thanks{\texttt{ro\_diff@yahoo.com}}}
\author[1]{Dorin Marinescu \thanks{\texttt{marinescu.dorin@ismma.ro}}}
\author[1,2]{Stefan-Gicu Cruceanu\thanks{\texttt{stefan.cruceanu@ismma.ro}}}
\affil[1]{``Gheorghe Mihoc-Caius Iacob'' Institute of Mathematical Statistics and Applied Mathematics of the ROMANIAN ACADEMY, Calea 13 Septembrie No. 13, PO Box 1-24, 050711 Bucharest, Romania}
\affil[2]{Corresponding author}

\hypersetup
{
  pdftitle={ASTERIX: Module for modelling the water flow on vegetated hillslopes},
  pdfsubject={Applied Mathematics},
  pdfauthor={Stelian Ion, Dorin Marinescu, Stefan-Gicu Cruceanu},
  pdfkeywords={Saint-Venant model, porosity, hexagonal raster, numerical scheme, hydrographic basin, C programming, open source}
}

\begin{document}


  {\def\thefootnote{}\footnotetext{\textit{Abbreviations:} PDE, Partial Differential Equation; ODE, Ordinary Differential Equation; FVM, Finite Volume Method; SDL, Simple DirectMedia Layer; GPL, GNU General Public License; GIS, Geographic Information System.}}

  \maketitle

  \begin{abstract}
    The paper presents an open source software for numerical
    integration of an extended Saint-Venant model used as a
    mathematical tool to simulate the water flow from
    laboratory up to large-scale spatial domains applying
    physically-based principles of fluid mechanics.  Many
    in-situ observations have shown that vegetation plays a
    key role in controlling the hydrological flux at
    catchment scale.  In case of heavy rains, the
    infiltration and interception processes cease quickly,
    the remaining rainfall gives rise to the Hortonian
    overland flow and the flash flood is thus initiated.  In
    this context, we also address the following problem: how
    do the gradient of soil surface and the vegetation
    influence the water dynamics in the Hortonian flow?  The
    mathematical model and ASTERIX were kept as simple as
    possible in order to be accessible to a wide range of
    stakeholders interested in understanding the complex
    processes behind the water flow on hillslopes covered by
    plants.
  \end{abstract}



  {\bfseries \emph{ACM:}} G.1.8; G.4

\section*{Software availability}
\label{software}
\noindent
Name of software: ASTERIX - Water Flow Module\\
Developers: Stelian Ion, Dorin Marinescu, Stefan G. Cruceanu.\\
Email: stefan.cruceanu@ismma.ro\\
Year first available: 2024.\\
Program Language: C.\\
Software Requirements: 32 or 64 bit Linux operating system, gcc
compiler, SDL libraries.\\
Cost: free under \href{https://www.gnu.org/licenses/gpl-3.0.en.html}{\textcolor{blue}{GNU license}}\\
Program size: 5 MB.\\
Availability: \url{http://www.ima.ro/software/ASTERIX\_flow\_2D.tar.xz}

\section{Introduction}
\label{section_intro}
\begin{figure}[!htbp]
  \centering
  \includegraphics[width=0.9\textwidth]{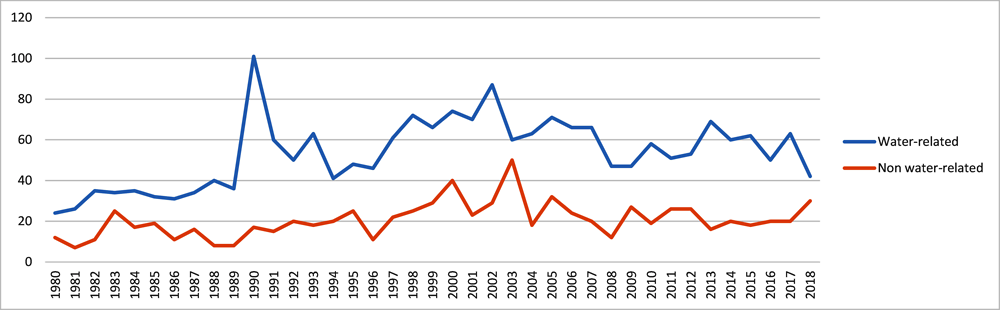}
  \caption{Number of water-related and non-water-related
    disasters in OECD countries.  Source: EM-DAT; The
    OFDA/CRED International Disaster Database, -
    Universit{\'e} catholique de Louvain (UCL) - CRED,
    Debarati Guha-Sapir - www.emdat.be, Brussels, Belgium.}
  \label{fig_disasters}
\end{figure}

Water is an essential element for life on earth, for plants,
animals, human beings.  Our daily comfort depends on water
resources and our way of life is mainly shaped by them, too.
The list of disasters over the past $50$ years is dominated
by the water-related ones which account for $70\%$ of all
deaths caused by natural catastrophes
\cite{worldbank_water_res_manag}.  Climate change is
affecting the hydrological cycle and increasing the
frequency and intensity of the storms.  Over $90\%$ of
disasters are weather-related, including drought and
aridification, wildfire, pollution and floods
\cite{unep_disasters_and_climate_change}.
Fig.~\ref{fig_disasters} illustrates a report regarding the
water-related and non-water-related disasters in OECD
countries \cite{oecd_book}.
Given all that, it is mandatory for all of us to understand
the hydrological cycles to properly manage the water
resources in order to protect our life and material
well-being.

The water distribution in a hydrographic basin is strongly
influenced by many factors related to land cover, land use,
soil type or soil surface gradients, \cite{land_cover_01,
  land_cover_02}.  Some of these factors play a key role in
controlling the destructive effects of the water-related
natural hazards.  Mathematical modelling of the hydrological
processes is an effective way to estimate the risks
associated to these hazards.  By modelling, one can make
scenarios, can evaluate the level of the risk exposure, and
can provide important information regarding an economical
water use (e.g. to agronomists).

\begin{figure}[!htbp]
  \centering
  \includegraphics[width=0.56\textwidth]{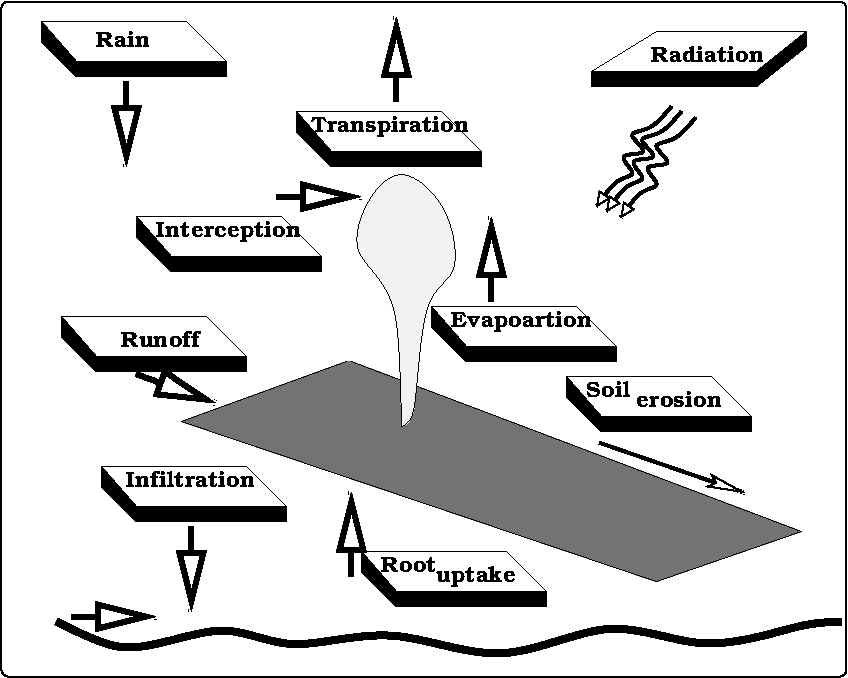}
  \caption{Watershed process}
  \label{fig_wathershed_process}
\end{figure}

Modelling the hydrological phenomenon is a very challenging
and difficult task with high complexity coming from the
various processes involved in: surface runoff, soil erosion,
precipitation, infiltration, evaporation, plant
transpiration, root uptake, etc., see
Fig.~\ref{fig_wathershed_process}.
We can add the multitude of the factors (only a few being
quantifiable) and the environmental heterogeneity (diversity
of vegetation and soil, variations in altitude, curvature of
the soil surface, etc.) and it becomes obvious that we
cannot take everything into account in one single
mathematical model.  Higher complexity does not guarantee an
overall higher performance; the performance can depend on
the target purpose or hydrological variables, \cite{orth}.
Besides this, a complex model often leads to problems of
over-parameterization and equifinality \cite{schoups2008,
  beven2006}.  By all means, we do not argue against complex
models, we only stress the fact that the modeller must be
careful about the processes to be included in the models.
{\it The best models are the ones which give results closest
  to reality and at the same time require least number of
  parameters and model complexity}, \cite{wheater_book}.  We
add that a good mathematical model must be {\it solvable}
and {\it physically relevant}.

The existent models cover a large spectrum of applications:
rainfall-runoff, flood hazard management, dam break
problems, agricultural hydrology, ecohydrology, and so on.
There are many classification criteria of the models and a
model can cast into one or more classes, depending on its
task, scale or structure.  For example, following
\cite{wheater_book}, the models can be classified as metric,
conceptual and physics-based ones.  Another classification
is considered in \cite{epa_rainfall_runoff} as empirical,
conceptual or physical.  The continuous representation of
hydrological processes is the main key to a physical model
whose equations are based on real hydrological responses.
The main advantages of such a model are:

$\bullet$ the spatial and temporal representations of the
variables can be accomplished at very fine scales;

$\bullet$ it can be applied to ungauged catchments, provided
that one can determine the parameters a priori;

$\bullet$ the effects of a catchment change can be
explicitly represented, \cite{wheater_book,
  epa_rainfall_runoff}.

\noindent In addition, there are many software products that
implement mathematical models, some of them are commercial
and some others are free.  From the last category, we can
mention \cite{R_source, Py_source, hype} for
hydrology-related {R} and Python tools and resources.  Among
the most known models in the class of physical models, we
mention SHE \cite{abbott_SHE}, MIKE-SHE \cite{mike_she},
KINEROS \cite{woolhiser_kineros}, VIC \cite{liang_VIC}, PRMS
\cite{singh_book}, SWASHES \cite{delestre}.

ASTERIX addresses the rainfall-runoff hydrological process
using the Saint-Venant equations as a mathematical model.
This software is mainly designed to study the combined
effects induced by the presence of plant on soil surface and
the variation of the gradient of the soil surface on the
water dynamics.  The plant cover and soil surface are
modelled with two space distributed functions: the porosity
and the soil altitude.  The water-plant stem and water-soil
frictional forces are quantified by two functions, both
proportional with the square of the water velocity.  The
coefficients of proportionality are of parameter function
type specific to plant cover and soil roughness,
respectively.  The soil altitude sets the geometrical
configuration where the studied processes take place, while
the porosity function and the frictional coefficients set
the internal structure of this geometrical configuration.
The space geometry and the internal structure give rise to
the environmental configuration.

The novelty of adding a pointwise distributed porosity
function to model the presence of plant cover allows the
study of water dynamics for various scenarios with different
vegetation distributions.

ASTERIX can be used from a laboratory level up to a
hydrographic basin scale.  One can use it for practical as
well as for theoretical applications, such as:

$\succ$ problems with water distribution: determination of
dry or flooded regions,

$\succ$ studies on the influence of plants on surface
runoffs, estimating their role in the continuum
{\bf S}oil-{\bf P}lant-{\bf A}tmosphere,

$\succ$ propagation of floods produced by torrential rains,

$\succ$ the effects of severe floods due to deforestation,

$\succ$ studies on Dam Break Problems,

$\succ$ studies on Riemann Problems, etc.

For practical applications in environmental problems, the
numerical approach chosen to solve such porous shallow-water
equations must come with a good balance between the
computational effort and the accuracy of the obtained
solution on one hand, and between the precision of the
measured data and the numerical accuracy on the other hand.
Although high-order schemes provide better accuracy, they
require higher computing effort.  Furthermore, using such
methods at the scale of hydrographic basins becomes very
difficult due to the excessive volume of processed data.
These are the reasons we developed and studied in
\cite{sds_apnum} a simple discrete model based on a
first-order numerical scheme with low algebraic calculations
and reduced memory requirements.  This scheme represents the
main module of our software ASTERIX for modelling the water
flow on vegetated surfaces.

The numerical scheme of the model behind this software was
subjected to several internal and external validation tests.
Tests such as steady flow over a bump, the oscillating
solution of Thacker's Problem, and the solution to the
Riemann Problem of the shallow-water equations were used for
the internal validation which consists in comparing the
numerical results with the exact solutions of the model.
For the external validation, we performed tests to compare
the numerical results with laboratory measured data
\cite{cadam, Dupuis2016}: dam break flow over a bump and in
a L-shaped channel, downslope flow through rigid vegetation.
A good agreement between the numerical and the
exact/measured data was obtained in all these cases.  A
qualitative analysis emphasizing the similarity between the
flow evolution on vegetated soil given by our numerical
model and the behavior of the observed phenomenon was also
performed to strengthen to confidence in the model.

We mention that the numerical modelling of the water flow in
ASTERIX is accomplished on hexagonal rasters due to the cell
adjacency and the local symmetry of the hexagonal structure.
The advantages of using this type of network are detailed in
\cite{sds-ADataPortingTool}.

This paper is organized as follows: in
Section~\ref{section_mmns} we present the PDE system of the
Saint-Venant model with vegetation and the numerical scheme
built to approximate this model and implemented in ASTERIX.
The software is described in Section~\ref{section_sd} and
some numerical applications add new examples for additional
validation on the method and the software used in
Section~\ref{section_ne}. The last section presents final
remarks and conclusions.

\section{Mathematical model and numerical scheme}
\label{section_mmns}
The water flow module of ASTERIX implements a numerical
scheme to approximate the Saint-Venant equations with
porosity.  In order to better understand this software, we
briefly present the key concepts concerning the mathematical
model and numerical solutions in what follows.  The reader
is referred to \cite{sds_apnum} for more details.

\subsection{Mathematical model}
\label{subsection_mm}
The Saint-Venant model with porosity we consider in this
paper is based on mass and momentum balance equations and
takes into account the variation of the soil surface, the
plant cover, and the rain as water source.  It reads as
  \begin{equation}
    \begin{split}
      \partial_t(\theta h) + \partial_1(\theta h v_1) + \partial_2(\theta h v_2)& = \mathfrak{M},\\
      \partial_t(\theta h v_a) + \partial_1(\theta h v_a v_1) + \partial_2(\theta h v_a v_2) + \theta h\partial_a{w} & = \mathfrak{t}^{p}_a + \mathfrak{t}^{s}_a, \quad a=1,2,
    \end{split}
    \label{swe_vegm_rm.02}
  \end{equation}
The unknown variables of this PDE system are
the water depth $h(t,\boldsymbol{x})$ and the two components
$v_1(t,\boldsymbol{x})$ and $v_2(t,\boldsymbol{x})$ of the
velocity $\boldsymbol{v}=(v_1,v_2)$.  The external data and
forcing functions of this model are:

(a) The altitude of the soil surface
\begin{equation}
  x^3=z(\boldsymbol{x}), \quad \boldsymbol{x}=(x^1,x^2)\in\Omega,
  \label{eq_soilsurf}
\end{equation}
where $\Omega\subset \mathbb{R}^2$ represents the horizontal
projection of the soil surface and is assumed to be a
connected bounded open set, $x^1$ and $x^2$ are the
coordinates along two orthogonal horizontal axes, and $x^3$
is the coordinate along the vertical axis.

(b) The porosity function
\begin{equation}
  \mbox{$\theta:\Omega\rightarrow [0,1]$}.
  \label{eq_theta}
\end{equation}
This function takes into account the presence of the cover
plant on the soil surface and is defined as the volume of
void space between the plant stems (volume which can be
filled with water) present in a unit volume.  We note that
$\theta = 1$ for bare soil and $\theta = 0$ for a complete
sealant plant cover.

(c) The water source.\\
The free term $\mathfrak{M}$ quantifies the contribution of
rain and infiltration to the water mass balance equation.

(d) The gravitational force.\\
Its action is quantified through
\begin{equation}
  w=g\left[z(\boldsymbol{x})+h\right],
  \label{eq_gforce}
\end{equation}
where $g$ represents the gravitational acceleration and
$z(\boldsymbol{x})+h$ is the free water surface level.

(e) The water-soil frictional forces \cite{rouse} are
quantified by
\begin{equation}
  \mathfrak{t}^{s}_a=-\theta \alpha_s (h) |\boldsymbol{v}|v_a,
  \label{eq_tsa}
\end{equation}
where $\alpha_s (h)$ is a non-negative function depending on
the given soil surface.  Experimental relations of Manning,
Ch{\'e}zy, or the Darcy-Weisbach are some of the most used
formulas for this coefficient in literature. The
Darcy-Weisbach expression
\begin{equation}
  \mathfrak{t}^{s}_a=-\theta \alpha_s |\boldsymbol{v}|v_a, \quad \rm{(Darcy-Weisbach),}
  \label{eq_tsa_darcy}
\end{equation}
has the advantage of being non-singular if the water depth
becomes zero.

(f) The resistance opposed by plants to the water flow
\cite{baptist, nepf} is quantified by
\begin{equation}
  \mathfrak{t}^{p}_a=-\alpha_p h \left(1-\theta\right)|\boldsymbol{v}|v_a,
  \label{eq_tpa}
\end{equation}
where $\alpha_p$ is a non-negative constant depending on the
type of vegetation.

The cumulative effects of the water-soil and water-plants
interactions is assumed to be additive in this model, and
one can thus introduce the resistance term
\begin{equation}
  \mathfrak{t}^{p}_a+\mathfrak{t}^{s}_a  := -{\cal K}(h,\theta)|\boldsymbol{v}|v_a,
  \label{eq_tau}
\end{equation}
where
\begin{equation}
  {\cal K}(h,\theta) = \alpha_p h \left(1-\theta\right) + \theta \alpha_s (h)
  \label{eq_K_h_theta}
\end{equation}
is the coefficient function of the frictional force of the
water-soil-plant system.

\begin{remark}
  The Saint-Venant model (\ref{swe_vegm_rm.02}) with
  porosity addresses the water flow on vegetated surfaces.
  The structure of the plant cover allows one to use spatial
  averaging methods, like in theory of flow transport in
  porous media, see \cite{Whitaker} for example.  In our
  case, the ``pore space'' is the interspace between the
  plant stems.  Urban runoff is another important domain
  where water flows through obstacles, in this case the
  ``pore space'' being the space between buildings.

  Apparently, the two domains are similar, but the structure
  of the pores is very different.  This structural
  difference imposes different mathematical models, and a
  model performing well in one case may be inadequate for
  the other case.

  There is a very rich literature devoted to the subject of
  urban runoff, see \cite{Guo2021} for a review of models in
  this area.  We bring also to the reader's attention the
  works \cite{Sanders2008, Guinot2017DualIP,
    SoaresFrazo2018, Varra2024} where one can find very
  important ideas and facts concerning the urban surface
  water flow models.
\end{remark}

\subsection{Numerical scheme}
\label{subsection_ns}

The numerical scheme introduced in \cite{sds_apnum} to
approximate the solution of model (\ref{swe_vegm_rm.02})
casts into the general class of method of lines.  We firstly
construct a discretization of the spatial variables and
differential operators only, and leave the time variable
continuous.  Then, we define a time discretization scheme to
integrate the ODE system previously obtained.  A first-order
FVM is used for the spatial discretization and a fractional
step method for the time discretization \cite{strang,
  veque-phd}.

The FVM method has three main ingredients:

(a) the net of the finite volume elements or the cells
decomposition of the fluid flow domain $\Omega$,

(b) the cell discretization of the variables (unknowns or
external data),

(c) the approximation of the gradient of the unknown
functions on the interfaces of two neighboring cells.

\medskip
\noindent (a) {\bf Cell decomposition}.

Let $\{\omega_i\}_{i=\overline{1,N}}$ be an admissible
polygonal partition \cite{veque} of $\Omega$,
\begin{equation}
  \Omega=\displaystyle\bigcup\limits_{i=1}^N \omega_i,
\end{equation}
with $\sigma_i$ being the area of the cell $\omega_i$.  For
any given cell $\omega_i$, we denote
\begin{itemize}
\item ${\cal N}(i)$ to be the set of all cell-indexes $j$
  for which the cell $\omega_j$ has a common side $(i,j)$
  with $\omega_i$,
\item $\boldsymbol{n}=(n_1,n_2)^T$ to be the unit normal
  vector pointing towards the outside of the boundary
  $\partial\omega_i$ of $\omega_i$.
\end{itemize}

\noindent (b) {\bf Cell discretization of the variables}.

Any function $\psi$ on a cell $\omega_i$ is approximated by
a constant value:
\begin{equation}
  \psi_i=\left.\psi(\boldsymbol{x})\right|_{\boldsymbol{x}\in \omega_i}.
\end{equation}

\noindent (c) {\bf Interface approximations}.

For any common interface $(i,j)$ between the cells
$\omega_i$ and $\omega_j$, we define the following
approximations:
\begin{equation}
  \theta h_{(i,j)}:=
  \left\{
    \begin{array}{ll}
      \theta_i h_i, & {\rm if}\; (v_n)_{(i,j)}>0,\\
      \theta_j h_j, & {\rm if}\; (v_n)_{(i,j)}<0,
    \end{array}
  \right.
  \label{fvm_2D_eq.06}
\end{equation}
\begin{equation}
  \widehat{\theta h}_{(i,j)} :=
  \left\{
    \begin{array}{ll}
      \theta h_{(i,j)}, & {\rm if}\; (v_n)_{(i,j)}\neq 0,\\
      \theta_i h_i, & {\rm if}\; (v_n)_{(i,j)}=0 \;{\rm and}\; w_i>w_j,\\
      \theta_j h_j, & {\rm if}\; (v_n)_{(i,j)}=0 \;{\rm and}\; w_i\leq w_j,
    \end{array}
  \right.
  \label{fvm_2D_eq.05}
\end{equation}
and
\begin{equation}
  \begin{array}{l}
    (v_a)_{(i,j)}:=\displaystyle\frac{v_{a\,i}+v_{a\,j}}{2}, \quad a=1,2,\\
    (v_n)_{(i,j)}:=\boldsymbol{v}_{(i,j)}\cdot\boldsymbol{n}_{(i,j)},
  \end{array}
  \label{fvm_2D_eq.04}
\end{equation}
with $\boldsymbol{n}_{(i,j)}$ denoting the unitary normal to
the common side of $\omega_i$ and $\omega_j$ pointing
towards $\omega_j$.

Given the particular (flux conservative - partial) form of
(\ref{swe_vegm_rm.02}), we now introduce the following
discrete operators:
\begin{equation}
  \begin{split}
    {\cal J}_{a\,i}(h,\boldsymbol{v}):=&-\displaystyle\sum\limits_{j\in{\cal N}(i)}l_{(i,j)}\theta h_{(i,j)} (v_a)_{(i,j)} (v_n)_{(i,j)},\\
    {\cal S}_{a\,i}(h,w):=&-\displaystyle\frac{1}{2}\sum\limits_{j\in{\cal N}(i)}l_{(i,j)}(w_j-w_i){\widehat{\theta h}_{(i,j)}} n_a|_{(i,j)},\\
    {\cal L}_i((h,\boldsymbol{v})):=&-\displaystyle\sum\limits_{j\in{\cal N}(i)}l_{(i,j)}\theta h_{(i,j)} (v_n)_{(i,j)},
  \end{split}
  \label{fvm_2D_eq_frac.01}
\end{equation}
where $l_{(i,j)}$ is the length of the common interface
$(i,j)$ between the cells $\omega_i$ and $\omega_j$.
${\cal S}_{a\,i}$ refers to the gradient of the free water
surface, while ${\cal J}_{a\,i}$ and ${\cal L}_i$ refer to
the flux of the linear momentum and mass, respectively.
Combining the spatial discretization with the time
fractional step method gives the full discrete scheme
associated to (\ref{swe_vegm_rm.02}) of the form
  \begin{equation}
    \begin{split}
      \sigma_i(\theta_i h_i)^{n+1}= 
      &\,\sigma_i(\theta_i h_i)^n+
      \triangle t_n{\cal L}_i((h,\boldsymbol{v})^{n})+
      \sigma_i\triangle t_n\mathfrak{M}_i(t^{n+1},h^{n+1}),\\
      \sigma_i(\theta_i h_iv_{a\,i})^{n+1}=
      &\,\sigma_i(\theta_i h_iv_{a\,i})^{n}+
      \triangle t_n\left({\cal J}_{a\,i}((h,\boldsymbol{v})^{n})+{\cal S}_{a\,i}((h,w)^{n})\right)-
      \triangle t_n\sigma_i{\cal K}_i(h^{n+1})|\boldsymbol{v}_i^{n+1}| v^{n+1}_{a\,i}, \quad a=1,2,
    \end{split}
    \label{fvm_2D_eq_frac.07}
  \end{equation}

\noindent
where a generic $\psi_i$ represents the value of $\psi$ on
$\omega_i$ for $i\in \overline{1,N}$, $\varphi^n$ is the
value of $\varphi$ at the moment of time $t^n$,
$\triangle t_n :=t^{n+1}-t^n$ is the time step, $|\cdot|$ is
the euclidean norm, and
\begin{equation}
  \mathfrak{M}_i(t,h):=\mathfrak{M}(t,h_i), \quad \quad
  {\cal K}_i(h):={\cal K}(h_i,\theta_i).
\end{equation}

To advance a time step, one must solve the system
(\ref{fvm_2D_eq_frac.07}) for each cell $\omega_i$, which
basically reduces to determining the water depth $h_i^{n+1}$
and velocity $\boldsymbol{v}_i^{n+1}$ at the time $t^{n+1}$
based on the known values $h_i^{n}$, $\boldsymbol{v}_i^{n}$,
$t^n$ and $\triangle t_n$.  The reader is referred to
Appendix~\ref{appendix_solve_hv} for extended details on
solving (\ref{fvm_2D_eq_frac.07}) and for details about the
restrictions on the time step $\triangle t_n$.

\section{Software description}
\label{section_sd}
ASTERIX is a software developed in {C} on a Linux operating
system, being dedicated to the simulation of water flow on
vegetated slopes.  It is based on the previously described
numerical scheme for the mathematical model
(\ref{swe_vegm_rm.02}) and uses a regular hexagonal network
for the spatial discretization.  SDL libraries are required
for visualizing the time distribution of water layer depth
or velocities.  ASTERIX is distributed under GPL free
software license:
\url{https://www.gnu.org/licenses/gpl-3.0.en.html}.

\begin{figure}[!htbp]
  \centering
  \includegraphics[width=0.7\textwidth]{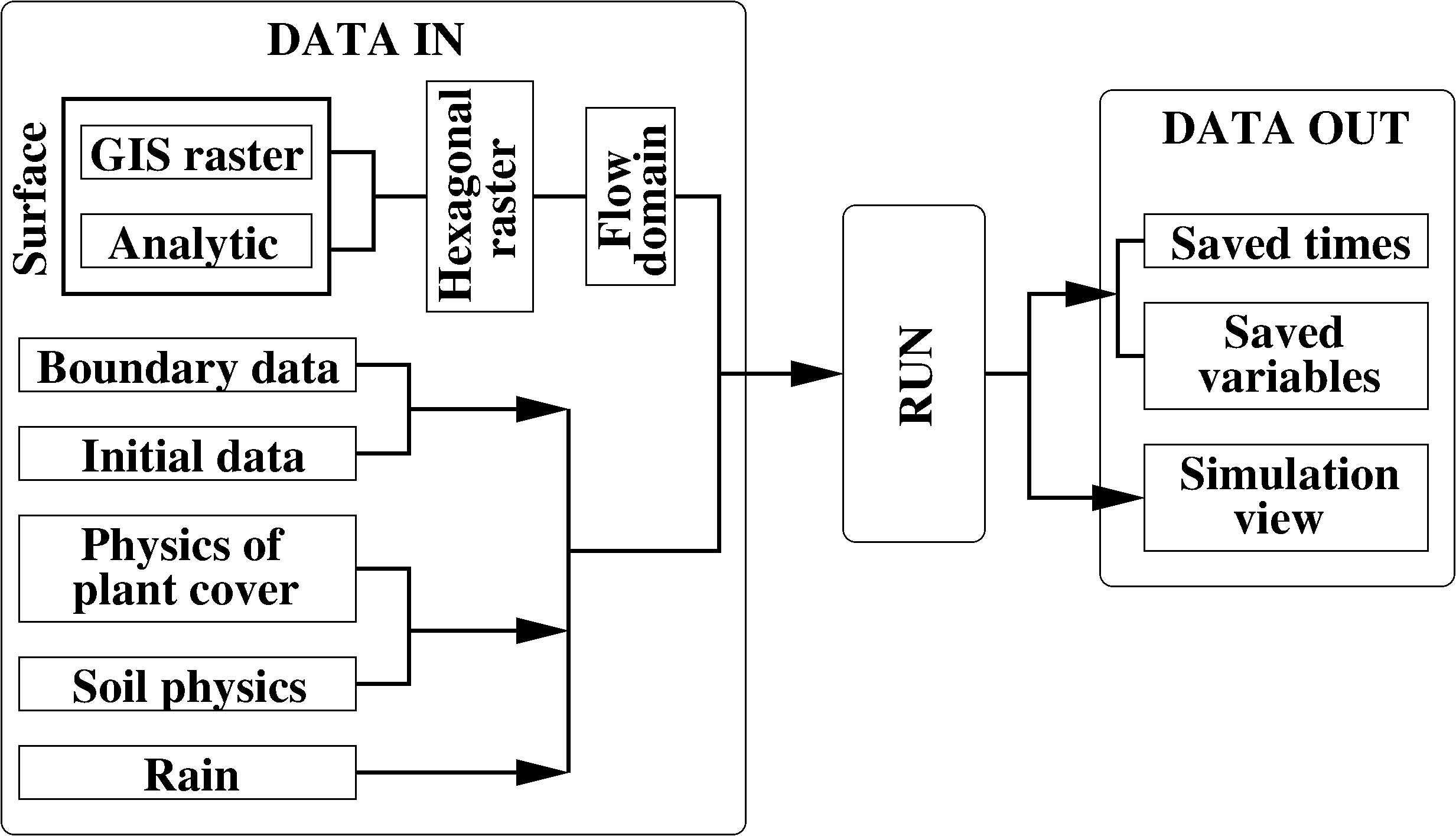}
  \caption{Diagram (sketch) of the ASTERIX software}
  \label{fig_Asterix_LogicalSchema}
\end{figure}

The source code of ASTERIX is a collection of files, most of
them containing many specific functions.  These component
functions can be logically grouped into three categories:
Data In, Running and Data Out.  One can improve the
performance or add new actions by modifying the existent
functions or adding new ones.  These categories are
associated to the three steps of the workflow (see
Fig.~\ref{fig_Asterix_LogicalSchema}) for operating the
software: i) pre-processing, ii) numerical solution, iii)
post-processing, respectively.  These three steps are
relatively independent, depending on the final purposes.
The first step is totally independent from the others.  The
second step depends on the output of the first step, but the
simulation can be redone by modifying only some parameters
specific to the numerical scheme (e.g maximal time step,
final time, etc).  The post-processing step depends on the
output of the second step.  These output data can be also
processed by other software.  We believe the source code and
the accompanying User Guide contain sufficient details to
allow a C advised programmer to make the desired
modifications.

\medskip \noindent {\bf Data In.}

The functions in this category collect the external
information and set values to the internal variables and to
the hyper-parameters of the program.  The most important
work is to process the environmental data concerning the
soil altitude, plant cover density, and plant and soil
characteristic parameters related to the frictional forces.

{\it Domain geometry and the cell partition of the flow
  domain}.  The topographic information of the soil provided
by a GIS raster is ported to a regular hexagonal raster
along this step.  One edits the file hexa\_parameters.dat in
order to establish the number of hexagonal cells on the
first row of the hexagonal grid.  The number of lines of the
hexagonal raster is then automatically calculated depending
on the geometry of the GIS raster.

Note: The program also allows the option to work on a
rectangular subdomain cut from the GIS raster.

The executable ``convert'' generates data files with
information on the hexagonal raster (e.g. cell radius, cell
centers, cell elevations, cell neighbors, cell interfaces,
cell types\footnote{In terms of belonging to the domain
  $\Omega$, the hexagonal raster of the soil surface
  contains internal cells and boundary cells, corresponding
  to the interior and boundary parts of $\Omega$,
  respectively.}, etc.) that are required by the main
program for flow simulation.  Details about the mathematical
framework for porting data methods can be found in
\cite{sds-ADataPortingTool}.

One can also artificially create the topography of a soil
using some mathematical functions.  In this case, the
topographic information data can be directly generated on a
regular hexagonal raster using the executable ``generate''.

{\it Boundary conditions}.  The boundary conditions are
themselves a rather difficult challenge due to the
hyperbolic type of the system of equations.  Furthermore,
identifying the parts of the border on which different
conditions must be established represents another
difficulty.  The software allows various boundary conditions
as free discharge, given mass flux, impermeable wall, and
Dirichlet.  Depending on the studied problem, ASTERIX allows
combinations of such boundary conditions as can be seen on
the various tests considered in Section~\ref{section_ne}.
We point out that default boundary condition is free
discharge; for other types of boundary conditions, the user
must define the configuration of the boundary with the
specific boundary conditions for the problem one has to
solve.  In principle, the software allows spatial and
time-varying boundary conditions, but such facilities are
not yet implemented at this stage; they must be hard-coded
by the user.

{\it Parameters}.  The flow simulation program can be
controlled through a series of parameters whose values can
be modified in the "parameter\_control.dat", file such as:
\begin{enumerate}
  \item {\it environmental parameters}
    \begin{enumerate}
    \item water-soil friction coefficient;
    \item water-plant friction coefficient;
    \item plant density;
    \end{enumerate}
  \item {\it program hyper-parameters}
    \begin{enumerate}
    \item total running time;
    \item maximum time step;
    \item the type of representation;
    \item choosing an initial water layer or using one rain
      simulator;
    \item extraction of snapshots at different moments of
      time;
    \item control of color shades for representations;
    \item the output data files we want to get at the end.
    \end{enumerate}
\end{enumerate}

\medskip \noindent {\bf Running.}

The flow is simulated via the executable ``2D\_flow'' on the
regular hexagonal grid.  The program can be run with or
without graphical representation.  The option {\it without
  graphics} shortens the runtime for finding the numerical
output.

\medskip \noindent {\bf Data Out.}

The program is able to generate files with numerical data
for certain quantities of interest at different moments of
time.  In particular, one can generate data files with the
evolution of the water depth and velocities at certain
predetermined points on the terrain.

When run {\it with graphics}, the program allows three
graphical representations:
\begin{enumerate}
\item the relief only;
\item the evolution of the water depth distribution;
\item the evolution of the water velocity field.
\end{enumerate}
If the option ``relief only'' is selected, ASTERIX allows
the user to inspect the terrain in order to find some
information about the hexagonal cells of the raster as the
coordinates of their centers, the altitudes, etc.  This
information can be useful when one desires to establish
monitoring points.

\medskip
Our software includes the GIS files of some hydrographic
basins with the help of which one can quickly test how the
software works on real topographies.  One can easily add new
GIS data files for simulations on other topographies.  The
software has also several analytic functions implemented for
simulating the flow for theoretical or laboratory
experiments.  The user with some experience in C can add new
functions in the available sources.

The directory structure and the operating diagram of ASTERIX
are synthesized in
Table~\ref{table_directory_structure_asterix} and
Fig.~\ref{fig_Asterix_LogicalSchema}, respectively.
Extended details can be found inside the User Guide.
\begin{table}[h]
  \centering
  \caption{Folder structure of ASTERIX with the key files}
  \begin{tabular}{ |l|p{1.6cm}|p{9.5cm}| }
    \hline
    {\bf Folder} & {\bf Content}
    & {\bf Description} \\\hline
    GIS & .asc
    & DEM files containing the topography of some real
      terrains (e.g. ampoi, lipaia, paul, susita).\\\hline
    gis\_hexa\_conversion & .c, .h
    & The C sources and header files for porting
      the raster data of a file from GIS into a hexagonal
      raster.\\\hline
    generate\_artificial\_gis & $\rm{.c, \, .h,}$ .dat
    & {The source code and header files
      for generating the hexagonal raster of a soil
      surface defined by a mathematical function in
      the .dat file.}\\\hline
    ini & .dat
    & Files with input data.\\\hline
    main & .c, .h
    & The main code files for the flow simulation.\\\hline
    out &
    & A place where the output (.dat, .png) files will be saved.\\\hline
  \end{tabular}
  \label{table_directory_structure_asterix}
\end{table}

The variables and the parameters describing model
(\ref{swe_vegm_rm.02}) and their encoded names inside the
software are also sketched in
Table~\ref{table_main_vars_of_asterix}.
\begin{table}[h]
  \centering
  \caption{The variables and the main parameters of
    (\ref{swe_vegm_rm.02}) with (\ref{eq_tau}) and
    (\ref{eq_K_h_theta}), and their encoded names in
    ASTERIX}
  \begin{tabular}{ |m{0.15\textwidth}|m{0.27\textwidth}|m{0.15\textwidth}|m{0.25\textwidth}| }
    \cline{2-4}
    \multicolumn{1}{c|}{} & {\bf Encoded name in ASTERIX} & {\bf Var \& param.} & {\bf Refers to}\\\hline
    \multirow{2}{*}{Unknown}
       & water\_vector[$\cdot$] & $h(t,\boldsymbol{x})$ & \multirow{2}{*}{Water dynamics}\\
       & velocities[$\cdot$][$\cdot$] & $\boldsymbol{v}(t,\boldsymbol{x})$ & {}\\\hline
    \multirow{4}{0.15\textwidth}{External data}
       & elevation\_vector[$\cdot$] & $z(\boldsymbol{x})$ & {Soil topography}\\\cline{2-4}
       & theta\_veg[$\cdot$] & $\theta(\boldsymbol{x})$ & \multirow{2}{*}{Physics of plant cover}\\
       & alpha\_veg & $\alpha_p$ & {}\\\cline{2-4}
       & alpha\_terr & $\alpha_s$ & {Physics of soil}\\\hline
  \end{tabular}
  \label{table_main_vars_of_asterix}
\end{table}

Table~\ref{table_asterix_mainfiles} presents a list of the
main files for running ASTERIX.
\begin{landscape}\begin{table}
  \centering
  \caption{Main files for running ASTERIX. be - binary executable; bs - Bourne-Again shell
    script; df - data file}
  \begin{tabular}{ |l|p{0.6cm}|p{3.0cm}|p{13cm}| }
    \hline
    {\bf Main Files} & {\bf Type} & {\bf Placement} & {\bf Description} \\\hline
    convert & be & gis\_hexa\_conversion
    & It converts the data from a DEM file from 'GIS/' into
      a hexagonal raster and populates 'ini/'’ with the
      input .dat files necessary for the main program
      '2D\_flow'.\\\hline
    generate & be & generate\_artificial\_gis
    & It generates a hexagonal raster for a particular type
      of artificial relief selected from
      field\_function\_parameters.dat and populates 'ini/'’ with the
      input .dat files necessary for the main program
      '2D\_flow'.\\\hline
    2D\_flow & be & main
    & The main executable file for simulating the water flow.\\\hline
    artificial\_relief\_selection.sh & bs & main
    & It allows one to select the desired
      artificial relief from field\_function\_parameters.dat
      generates the hexagonal raster through the executable
      'generate'.\\\hline
    natural\_relief\_selection.sh & bs & main
    & It allows one to select the desired
      natural relief from 'GIS/' and generates the hexagonal
      raster through the executable 'convert'.\\\hline
    vpha.sh & bs & main
    & It allows one to choose the resolution of the
      hexagonal network associated by editing the file
      'hexa\_parameters.dat' and generates the hexagonal
      raster through the executable 'generate' for a
      specific artificial relief. \\\hline
    vphn.sh & bs & main
    & It allows one to choose the resolution of the
      hexagonal network by editing the file
      'hexa\_parameters.dat' and converts the data from a
      DEM file from 'GIS/' into a hexagonal raster through
      the executable 'convert' for a specific natural relief. \\\hline
    vpc.sh & bs & main
    & It allows one to quickly edit the file
      'control\_parameters.dat' and to modify the data from
      this file. \\\hline
    no\_graphics.sh & bs & main
    & It creates the folder 'main-no\_graphics/' and
      populates it with the simplified code files (without
      any graphics) of ASTERIX. The flow simulation with
      numerical data output only can now be run from this
      folder. \\\hline
    control\_parameters.dat & df & ini
    & It contains the values of different parameters
      necessary for running ASTERIX at a given
      resolution. \\\hline
    hexa\_parameters.dat & df & ini/ini\_hexa
    & It contains the values of the parameters for the
      hexagonal grid resolution. \\\hline
    field\_function\_parameters.dat & df & generate\_artificial\_gis
    & It contains the data for some types of artificial
      reliefs. \\\hline
    susita.asc & df & GIS
    & A DEM file with the GIS data for the Susita River
      basin. \\\hline
    ampoi.asc & df & GIS
    & A DEM file with the GIS data for the Ampoi's Valley. \\\hline
    lipaia.asc & df & GIS
    & A DEM file with the GIS data for the Lipaia's Valley. \\\hline
    paul.asc & df & GIS
    & A DEM file with the GIS data for the Paul's Valley. \\\hline
  \end{tabular}
  \label{table_asterix_mainfiles}
\end{table}\end{landscape}

Assuming the user has downloaded, decompressed the archive
of ASTERIX and compiled the files, we now present a simple
example for operating the software on \c{S}u\c{s}i\c{t}a
River basin described in
Subsection~\ref{sect_SimulationonSusita}.

Go to folder ’main/’ and execute the following commands:

\$ ./natural\_relief\_selection.sh susita

\$ ./2D\_flow

\noindent
The last command runs ASTERIX with a graphic representation
of the water depth over time;  the user can also run

\$ ./2D\_flow -r2

\noindent
for a graphic representation of velocity field over time.
More details on operating the software can be found in the
accompanying 'User Guide'.

\section{Performance Tests}
\label{section_ne}
A suite of examples is presented in \cite{sds_apnum} for the
purpose of validating the numerical scheme.  In what
follows, we add some new examples for testing the
performance of the software.  The numerical results are
compared with the analytic solutions for some of them and
with the laboratory measurements for others.  We analyze the
response of the software to real terrain data and perform a
sensitivity analysis with respect to some model as well as
some software parameters.

\subsection{Theoretical Tests}
\subsubsection{Flow on radial symmetric surfaces}
\begin{figure}[!htbp]
  \hspace{-3mm}
  \begin{tabular}{ ccc }
    {} & \hspace{0mm}\small{Crater type} & \hspace{0mm}\small{Hillock type} \\
    \begin{turn}{90}\hspace{6mm}\small{Soil surface}\end{turn}\hspace{-5mm}
       & \includegraphics[width=0.48\textwidth, height=2.6cm]{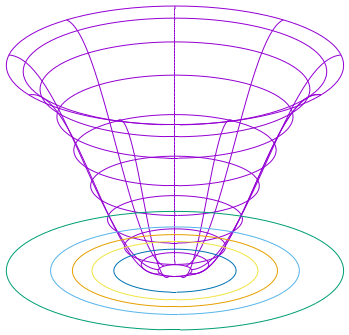}\hspace{-5mm}
       & \includegraphics[width=0.48\textwidth, height=2.6cm]{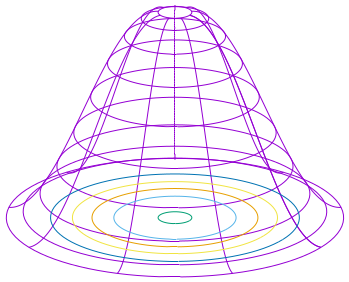}\\
    {} & {Radial section} & {Radial section}\\
    \begin{turn}{90}\hspace{6mm}\small{Soil surface}\end{turn}\hspace{-5mm}
       & \includegraphics[width=0.48\textwidth, height=2.6cm]{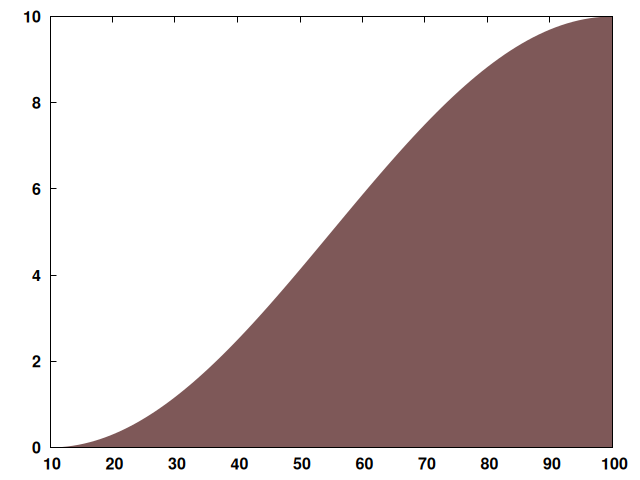}\hspace{-5mm}
       & \includegraphics[width=0.48\textwidth, height=2.6cm]{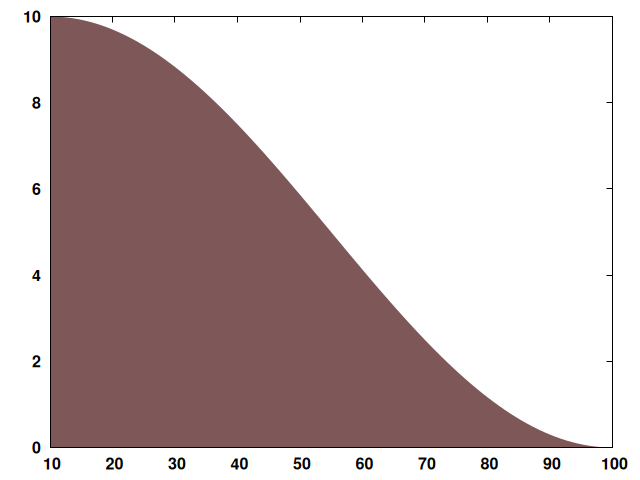}\\
    \begin{turn}{90}\hspace{2mm}\small{Water depth [m]}\end{turn}\hspace{-5mm}
       & \includegraphics[width=0.48\textwidth, height=2.6cm]{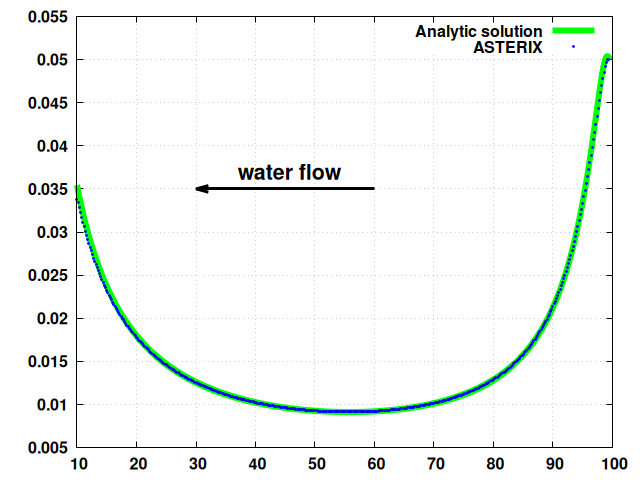}\hspace{-5mm}
       & \includegraphics[width=0.48\textwidth, height=2.6cm]{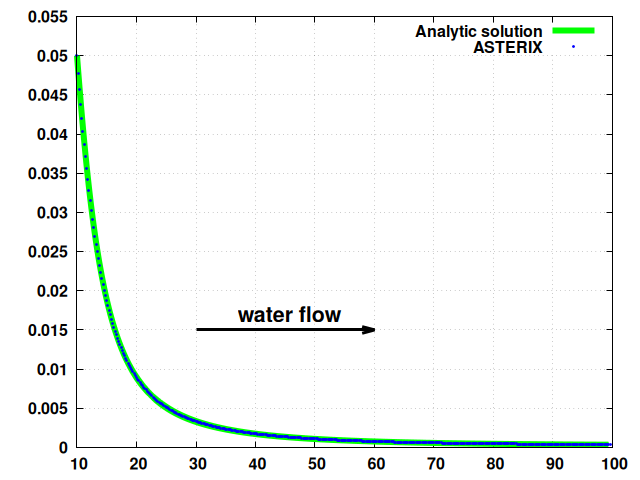}\\
    \begin{turn}{90}\hspace{2mm}\small{Water vel. [m/s]}\end{turn}\hspace{-5mm}
       & \includegraphics[width=0.48\textwidth, height=2.6cm]{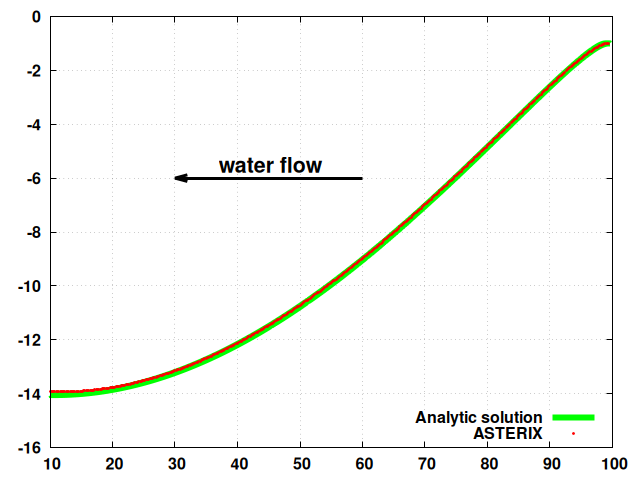}\hspace{-5mm}
       & \includegraphics[width=0.48\textwidth, height=2.6cm]{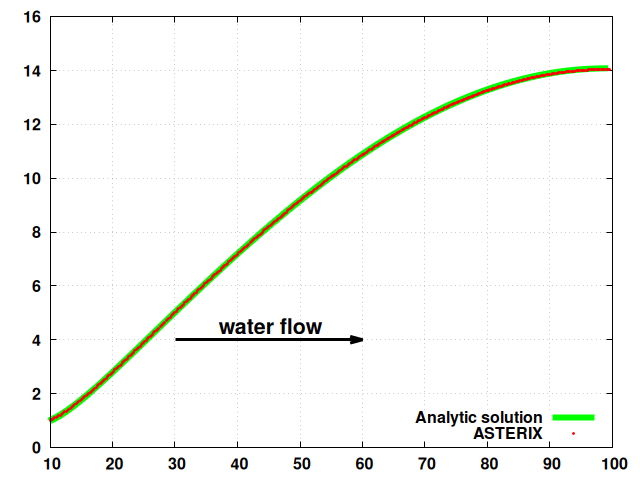}\\
    {} & \hspace{8mm}\small{$x$ [m]} & \hspace{8mm}\small{$x$ [m]}
  \end{tabular}
  \caption{The stationary solution of the water flow on two
    radial symmetric surfaces: crater type (first column) and
    hillock type (second column).  Boundary conditions:
    $h=0.05$ m and $v=1$ m/s at the top of the surface and
    free drainage at the bottom.}
  \label{fig_flow_on_radial_symm_surfs}
\end{figure}

A first test was performed on two theoretical surfaces with
radial symmetry.  These soil surfaces pictured on the first
row of Fig.~\ref{fig_flow_on_radial_symm_surfs} are built
using a parametric representation
\begin{equation}
  \left\{
    \begin{array}{l}
      x^1=r\cos{\theta}\\
      x^2=r\sin{\theta}\\
      x^3=f(r)
    \end{array}
  \right.
  \label{conic_surface}
\end{equation}
on a annulus
\begin{equation*}
  D:=\{(r,\theta)\left| \, r\in[r_0,r_1], \, \theta\in[0,2\pi)\right.\},
\end{equation*}
with $r_0=10$, $r_1=100$, and
\begin{equation}
  f(r) = \displaystyle A + s \cdot A \cos \left( \frac{\pi (r-r0)}{r_1-r_0} \right),
  \label{eq_function_fr}
\end{equation}
with $A=10$ m, $s=-1$ for the Crater surface on the left
column and $s=+1$ for the Hillock surface on the right
column.  The boundary conditions are of Dirichlet type for
the upper part and free discharge for the bottom part of
both surfaces:
\begin{equation}
  h(t,\boldsymbol{x}) = h_0, \quad \boldsymbol{v}(t,\boldsymbol{x})=-v_0\boldsymbol{n},
  \quad \forall\boldsymbol{x}\in\partial D^{\rm in},
  \label{numerics-bc}
\end{equation}
where $\partial D^{\rm in}$ is the upper boundary of the
surface, $\boldsymbol{n}$ is the external oriented unit
normal vector of $\partial D^{\rm in}$ at $\boldsymbol{x}$,
and \mbox{$h_0=0.05$ m}, \mbox{$v_0=1$ m/s}.  The advantage
of such radial symmetric surfaces is that we can write the
analytic solution of the flow in steady state
\cite{sds_atee2023}.

The numerical solution was obtained with ASTERIX on a
hexagonal raster with approximately $900000$ cells of radius
$0.115$ m.  A comparison between the analytic and numerical
water depth and velocity distributions of the stationary
solution of the water flow along a radial section is
presented in Fig.~\ref{fig_flow_on_radial_symm_surfs}.
For the errors between the two solutions we used the
formulas
\begin{equation*}
  \epsilon_h = \frac{1}{N} \sum_{i=1}^{N} {\left| \frac{h_n(r_i)-h_a(r_i)}{h_a(r_i)} \right|}, \quad\quad
  \epsilon_v = \frac{1}{N} \sum_{i=1}^{N} {\left| \frac{v_n(r_i)-v_a(r_i)}{v_a(r_i)} \right|},
\end{equation*}
where $N$ is the number of points $r_i$ from $D$ at which
the numerical solution is evaluated. $h_n$ and $h_a$ stand
for the numerical and analytic water depth, respectively.
$v_n$ and $v_a$ stand for the numerical and analytic water
velocity, respectively.

The small values of the errors presented in Table~\ref{table_err}
show there is a good agreement between the analytic and
numerical solutions.
\begin{table}[h]
  \centering
  \caption{Relative mean error between the numerical and analytic steady state solutions.}
  \begin{tabular}{ ccc }
    \toprule
    {} & {Crater type soil} & {Hillock type soil}\\
    \midrule
    $\epsilon_h$ & $0.0055$ & $0.0111$\\
    $\epsilon_v$ & $0.0040$ & $0.0018$\\
    \bottomrule
  \end{tabular}
  \label{table_err}
\end{table}

\subsubsection{Thacker's Problem}
The next 2D experiment, Thacker's Problem, is a powerful
test to check the ability of our software to catch and
highlight the propagation of a wet/dry front.  The
singularity appearing at the wet/dry contact between water
and soil surfaces raises difficulties for many numerical
methods, but a huge advantage of this problem is that it can
be exactly solved \cite{delestre, thacker, sampson,
  matskevich} and thus, one can make use of its analytic
solution to theoretically test the developed numerical
method.

Thacker's Problem is basically formulated as the PDE system
(\ref{swe_vegm_rm.02}) modelling the flow dynamics over a
paraboloid type surface
\begin{equation}
  z(x,y) = a (x-x_0)^2 + b (y-y_0)^2, \quad a,b>0,
  \label{eq_soil_surface_thacker}
\end{equation}
in the absence of vegetation ($\theta = 1$) and mass source
($\mathfrak{M} =0$), for a soil resistance term of the form
\begin{equation}
  \mathfrak{t}^{s}_a = -\tau h v_a, \quad a=1,2,
  \label{eq_soil_friction_thacker}
\end{equation}
where $\tau$ is a proportionality coefficient, and for a
fluid velocity which does not depend on the space variable.

Depending on the values of some parameters, the exact
solutions of this problem can be classified into
oscillating, non-oscillating and mixed ones.  In
\cite{sds_apnum}, we have considered the case of Thacker's
Problem with an oscillating solution and compared it with
the numerical one described there and implemented in
ASTERIX.  In this subsection, we present a comparison
between a non-oscillating solution for Thacker's Problem and
the numerical solution provided by ASTERIX\footnote{The soil
  resistance term $\mathfrak{t}^{s}_a$ is theoretically
  chosen to be proportional to the water depth and velocity
  in Thacker's Problem in order to obtain an analytic
  solution.  Note that this form
  (\ref{eq_soil_friction_thacker}) is different from
  (\ref{eq_tsa_darcy}), and ASTERIX was modified
  accordingly.}.  The non-oscillating solution was obtained
using $a = 1.25\cdot 10^{-3}$, $b = 5\cdot 10^{-3}$,
$x_0 = y_0 = 500\ {\rm m}$ in
(\ref{eq_soil_surface_thacker}), and a proportionality
coefficient of the friction force $\tau = 0.7\ {\rm s}^{-1}$
in (\ref{eq_soil_friction_thacker}).  The reader is referred
to Appendix~\ref{sect_apendix_thacker_nonoscilatoriu} for
extended mathematical details about the analytic solution to
this problem.

\begin{figure}[!htbp]
  \centering
  \begin{tabular}{cc}
    \includegraphics[width=0.45\linewidth]{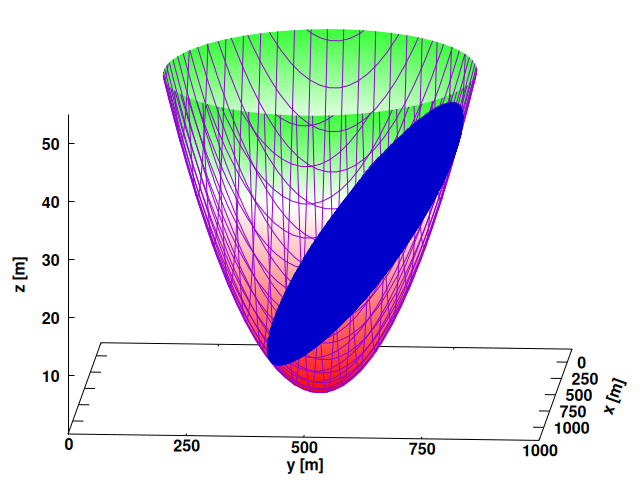}
    &\includegraphics[width=0.45\linewidth]{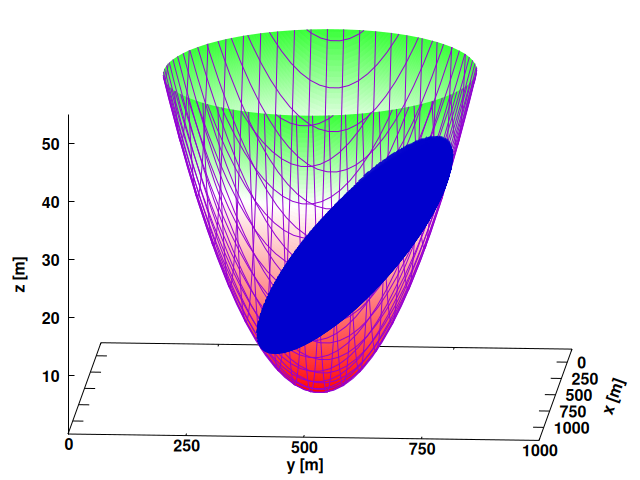}\\
    $t=10\,{\rm s}, \qquad err=0.0107$ & $t=30\,{\rm s}, \qquad err=0.0102$\\
    &\\
    \includegraphics[width=0.45\linewidth]{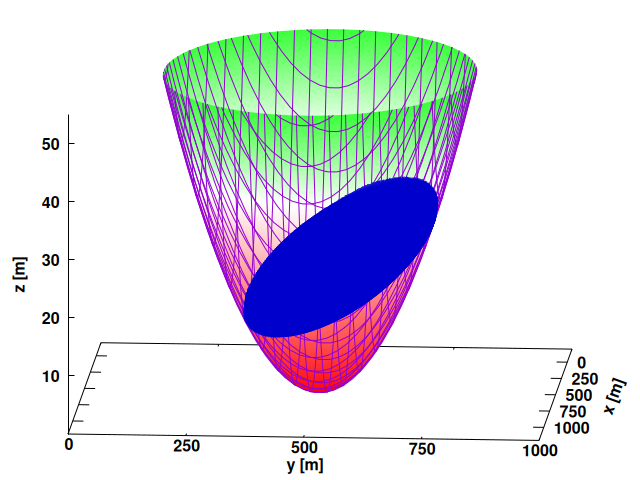}
    &\includegraphics[width=0.45\linewidth]{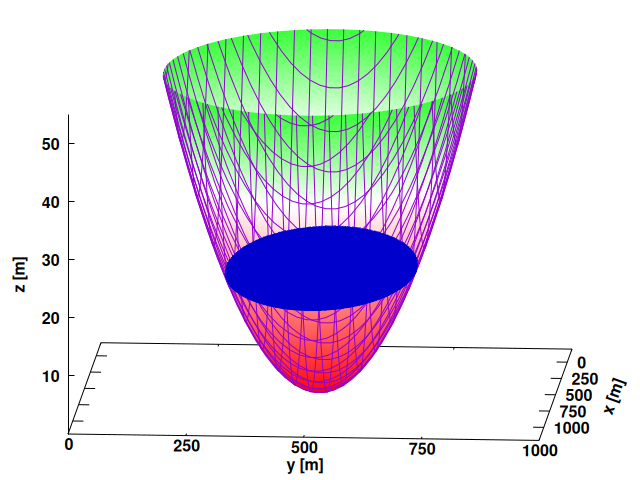}\\
    $t=70\,{\rm s}, \qquad err=0.0122$ & $t=330\,{\rm s}, \qquad err=0.0045$
  \end{tabular}
  \caption{2D Thacker's Problem: dynamics of the numerical
    free water surface.  These snapshots are obtained using
    Gnuplot on the data calculated with ASTERIX at four
    different moments of time:
    $t=10,\, 30,\, 70,\, 330\, {\rm s}$.  The error $err$
    between the numerical and exact values of the free water
    surface at each of the four moments of time is
    calculated with
    $err=||((h+z)^{app}-(h+z)^{ex})/(h+z)^{ex}||_{\infty}$.}
  \label{fig_thacker2D_2}
\end{figure}

We consider the following initial data:
\begin{equation}
  u_0=v_0=0, \quad u^{\prime}_0=0.02g, \quad v^{\prime}_0=-0.1g, \quad w_0 = 15\,{\rm m},
\end{equation}
where $g=9.81\ {\rm m}\cdot {\rm s}^{-2}$ is the
gravitational acceleration.  Fig.~\ref{fig_thacker2D_2}
shows that our numerical solution (obtained on a regular
hexagonal mesh with $487123$ cells of radius
$0.888231\, {\rm m}$) is in accordance with the
non-oscillating planar free water surface given by the
analytic solution to the Thacker's Problem.

\begin{figure}[!htbp]
  \centering
  \begin{tabular}{ ccc }
    \begin{turn}{90}\hspace{25mm}{Error}\end{turn}
       &\includegraphics[width=0.40\linewidth]{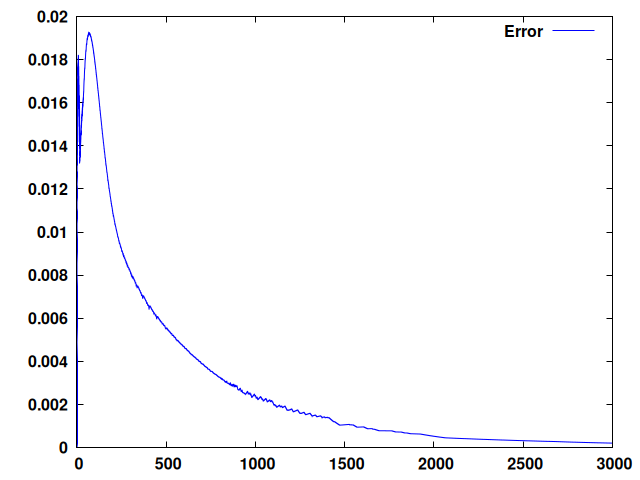}
       &\includegraphics[width=0.40\linewidth]{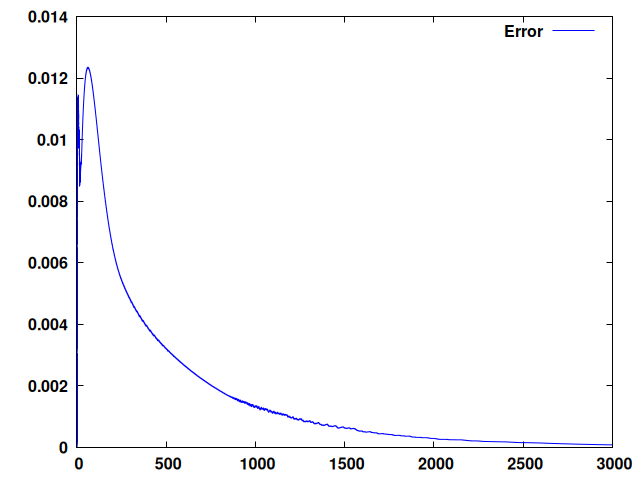}\\
    {} &time [s] & time [s]
  \end{tabular}
  \caption{2D Thacker's Problem: time evolution of the error
    $err=||((h+z)^{app}-(h+z)^{ex})/(h+z)^{ex}||_{\infty}$
    between the analytic and numerical solutions on two
    hexagonal networks with different cell sizes:
    $1.65\,{\rm m}$ and $0.888\,{\rm m}$ for the left and
    right picture, respectively.}
  \label{fig_thacker2D_1}
\end{figure}

The evolution of the error
$$err=||((h+z)^{app}-(h+z)^{ex})/(h+z)^{ex}||_{\infty}$$
between the exact and numerical solution is pictured in
Fig.~\ref{fig_thacker2D_1}.

\begin{figure}[!htbp]
  \centering
  \includegraphics[width=0.5\linewidth]{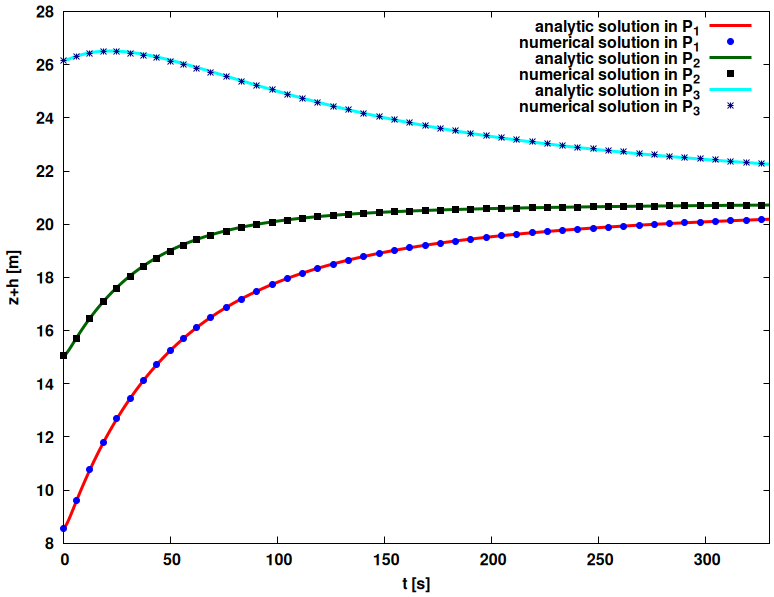}
  \caption{2D Thacker's Problem: time evolution of the free
    water surface at the three points from
    Table~\ref{table_3points}.}
  \label{fig_thacker_3points}
\end{figure}

Fig.~\ref{fig_thacker_3points} shows a comparison between
the time evolution over time of the numerical and analytic
solutions of the free water surface at the three different
points $P_1$, $P_2$, $P_3$ from Table~\ref{table_3points}.
\begin{table}[h]
  \centering
  \caption{Data for the three points in Tacker's Problem
    where the evolution over time of the free water surface
    is drawn in Fig.~\ref{fig_thacker_3points}.}
  \begin{tabular}{ c|ccc }
    \toprule
        & $P_1$ & $P_2$ & $P_3$\\
    \midrule
    $x$ & $577.692$ & $500.000$ & $264.615$\\
    $y$ & $451.221$ & $500.518$ & $564.471$\\
    $z$ & $1.944$   & $0$       & $9.004$\\
    \bottomrule
  \end{tabular}
  \label{table_3points}
\end{table}

\subsubsection{Test on a Riemann Problem with vegetation}
\label{sect_RP}

Riemann Problem is a mathematical subject associated to the
Dam Break Problem, a real physical phenomenon.  It
represents a very important issue in the theory of
hyperbolic systems because one can find an analytical
solution of the problem for a broad spectrum of initial
data.

The analytic solution is set up by a bunch of very special
solutions, namely, shock waves and simple waves.  The
combination of these waves in a packet is ruled by a
physical admissibility criteria.  In the absence of such
criteria the Riemann Problem can have a multitude of
solutions, some of them having no physical sense.  A
numerical scheme is considered to be at least physically
acceptable if it correctly predicts both types of waves.

The existence of the solution and its very special structure
make the Riemann Problem almost a mandatory test for any
numerical scheme dedicated to the shallow water equation.

The objective of our test is twofold: one aim is to verify
how well the numerical solution approximates the true
solution of the problem, and the other one is to verify if
the 2D numerical scheme furnishes a markedly 1D solution for
appropriate initial data and boundary conditions.

In this section, we choose a very difficult problem where
both the soil surface and the porosity function exhibit
simultaneous jumps.  The piecewise constant soil surface and
porosity functions
\begin{equation}
  (\theta,z)(x,y)=
  \left\{
    \begin{array}{ll}
      (\theta_L,z_L), & x \leq 9\\
      (\theta_R,z_R), & x > 9
    \end{array}
  \right. .
\end{equation}
are given on a $18$ m long and $1$ m wide horizontal
rectangular channel.  The piecewise constant initial data
are given by
\begin{equation}
  \left. h(x,y) \right|_{t=0}
  =\left\{
    \begin{array}{ll}
      h_L, & x \leq 9\\
      h_R, & x > 9
    \end{array}
  \right. ,
\end{equation}
\begin{equation}
  \left. \boldsymbol{v}(x,y) \right|_{t=0}
  =\left\{
    \begin{array}{ll}
      (u_L,0), & x \leq 9\\
      (u_R,0), & x > 9
    \end{array}
  \right. ,
\end{equation}
whose numerical values are in Table~\ref{table_DateRP}.
\begin{table}[h]
  \caption{Data for the 2D Riemann Problem}
  \centering
  \begin{tabular}{ cccccccc }
    \toprule
    $\theta_L$ & $z_L$ & $h_L$ & $u_L$   & $\theta_R$ & $z_R$ & $h_R$ & $u_R$\\
     -         & {[m]} & {[m]} & {[m/s]} &  -         & {[m]} & {[m]} & {[m/s]}\\
    \midrule
    $0.8$ & $1.0$ & $0.2$ & $5.00$ & $1.0$ & $1.2$ & $0.6$ & $1.33$\\
    \bottomrule
  \end{tabular}
  \label{table_DateRP}
\end{table}

\begin{figure}[!htbp]
  \centering
  \begin{tabular}{cc}
    \includegraphics[width=0.48\linewidth]{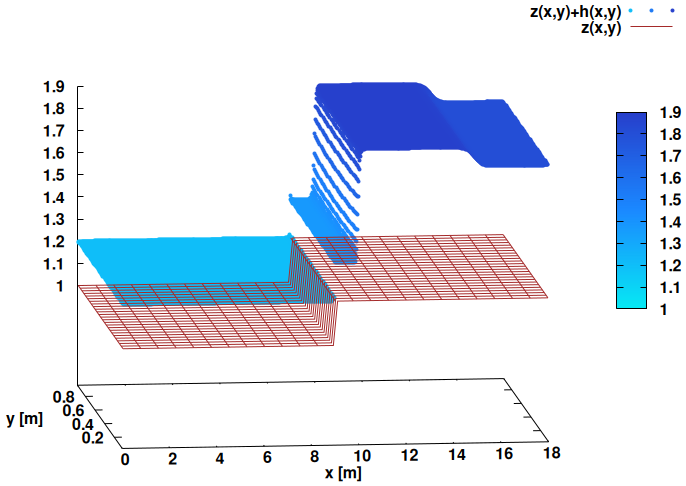}
    &\includegraphics[width=0.48\linewidth]{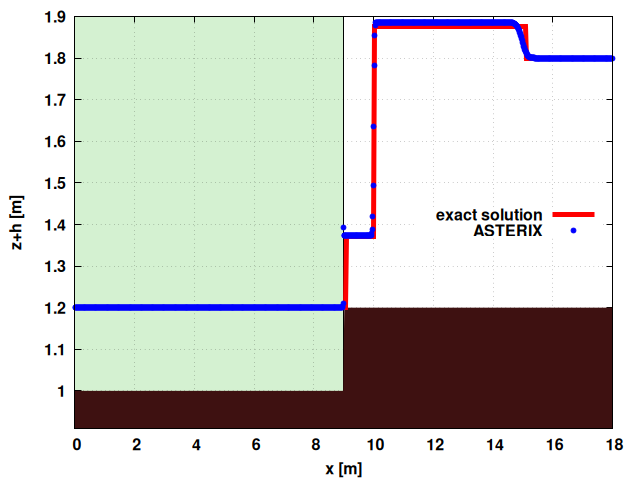}
  \end{tabular}
  \caption{2D Riemann's Problem: free water surface found
    with ASTERIX at $t=1.5\;{\rm s}$ (left picture) and a
    comparison between the analytic and numerical solutions
    from a longitudinal channel section (right picture,
    where the vegetated and the bare parts of the soil are
    marked with green and white, respectively).}
  \label{fig_Riemann2D}
\end{figure}

The exact solutions of Riemann Problem are difficult to find
and have only been studied in the one-dimensional case.  For
the 2D test considered in this subsection, the problem of
finding the solution can be reduced to a 1D case for which
it was shown the existence and uniqueness, \cite{smc-rp-19}.
The water surface calculated with ASTERIX at
$t=1.5\;{\rm s}$ and a comparison between the analytic
solution and the numerical data on a longitudinal section
are presented in Fig.~\ref{fig_Riemann2D}.

The value of the mean relative error
\begin{equation*}
  \frac{1}{N}\sum_{i=1}^N \left| \frac{(h+z)_i^{app}-(h+z)_i^{ex}}{(h+z)_i^{ex}} \right|
\end{equation*}
between the numerical ({\it app}) and exact ({\it ex})
values of the free water surface is 0.003785, where $N=1122$
is the number hexagonal centers along the longitudinal
section.

We remark that the three shock waves developed by the exact
solution are also caught by ASTERIX.  Furthermore, one can
observe that the hexagonal grid used by our software does
not induce numerical adverse effects such as additional head
losses due to the impossibility of the hexagons to align
exactly with the channel walls.

\subsection{Tests on real data}
\subsubsection{Simulation on \c{S}u\c{s}i\c{t}a River basin
  with vegetation}
\label{sect_SimulationonSusita}
\c{S}u\c{s}i\c{t}a's hydrographic basin is the catchment
area of \c{S}u\c{s}i\c{t}a River in Vrancea County, Romania.

\begin{figure}[!htbp]
  \centering
  \includegraphics[width=0.99\linewidth]{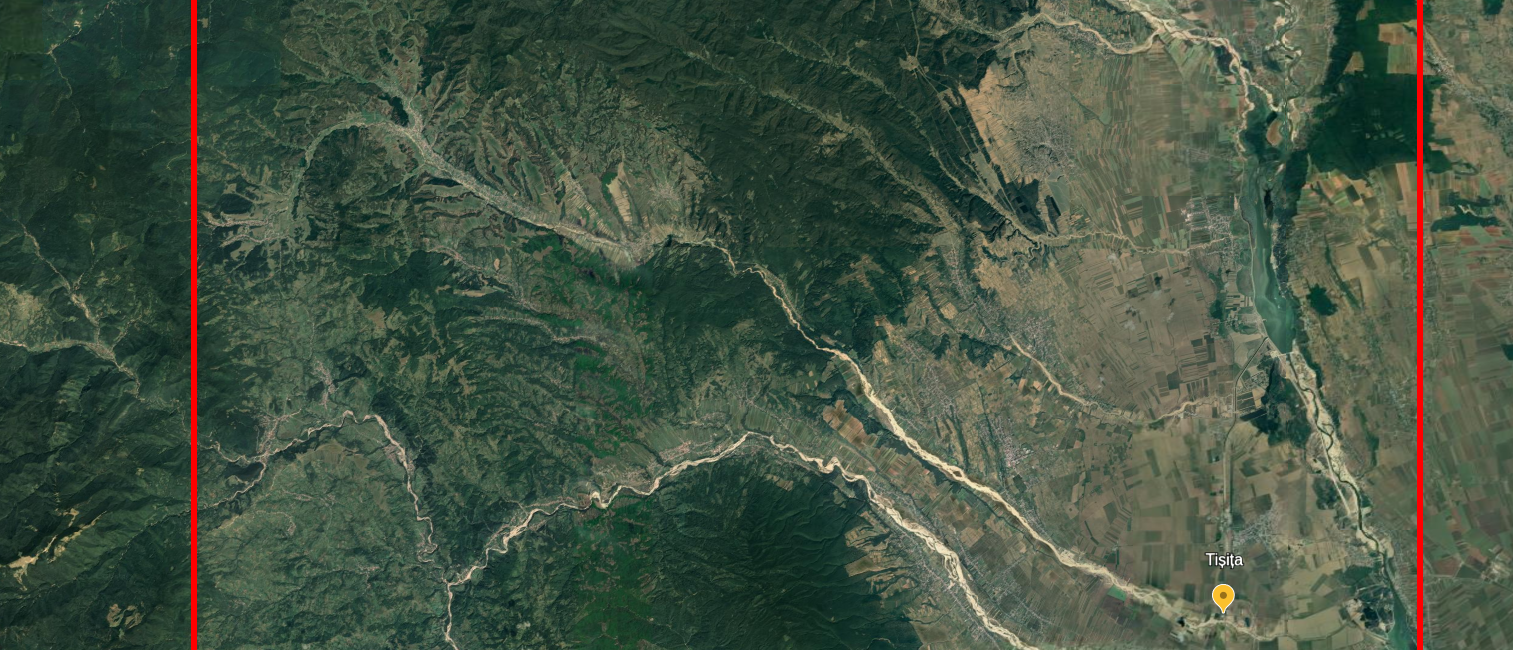}
  \\\vspace{0.5em}
  \hspace{18mm}\includegraphics[width=0.85\linewidth]{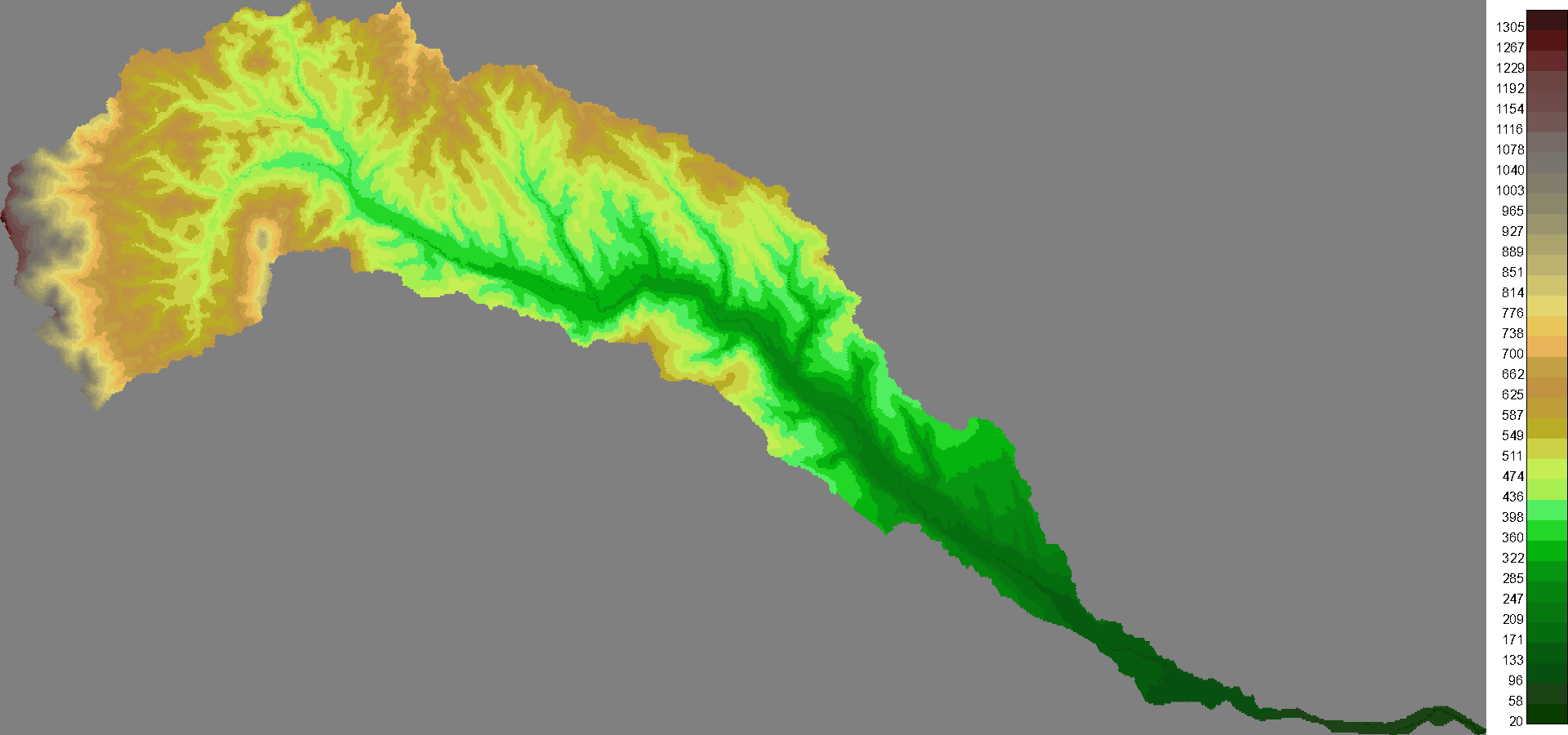}
  \caption{\c{S}u\c{s}i\c{t}a's hydrographic
    basin.  Satellite image \cite{google_earth_susita} and
    the terrain reconstructed (from GIS data) on a hexagonal
    network.}
  \label{fig_susita_foto}
\end{figure}

Unfortunately, we do not have data for the water
distribution, plant cover density and measured velocity
field in a hydrographic basin to compare our numerical
results with.  However, to be closer to reality, we have
used the GIS data for an approximately $370\, {\rm km}^2$
soil surface included in a $64 \times 28\, {\rm km}^2$
rectangular area.  More precisely, this rectangular area,
Fig.~\ref{fig_susita_foto}, lies north-west of
Ti\c{s}i\c{t}a whose geographical
coordinates are $45^{\circ}$ $50^{\prime}$
$36^{\prime\prime}$ North, $27^{\circ}$ $13^{\prime}$
$32^{\prime\prime}$ East.

We accomplished a theoretical experiment: starting with a
uniform rainfall generated by a triangular model (see
Appendix~\ref{sect_apendix_rainfall_triangular} for extended
details) and characterized by
\begin{itemize}
\item the total time $T_d = 1000\, {\rm s}$ of the rain,
\item the maximum intensity $i_a = 0.0732\, {\rm mm/s}$ of the rain, and
\item the time $t_a = 250\, {\rm s}$ at which this maximum
  intensity is recorded,
\end{itemize}
on the entire basin and using two different uniform
vegetation densities ($\theta_1 = 1.0$ and
$\theta_2 = 0.97$), we ran ASTERIX on a network with $82148$
internal hexagonal cells of radius $41\, {\rm m}$.

\begin{figure}[!b]
  \centering
  \includegraphics[width=0.95\textwidth]{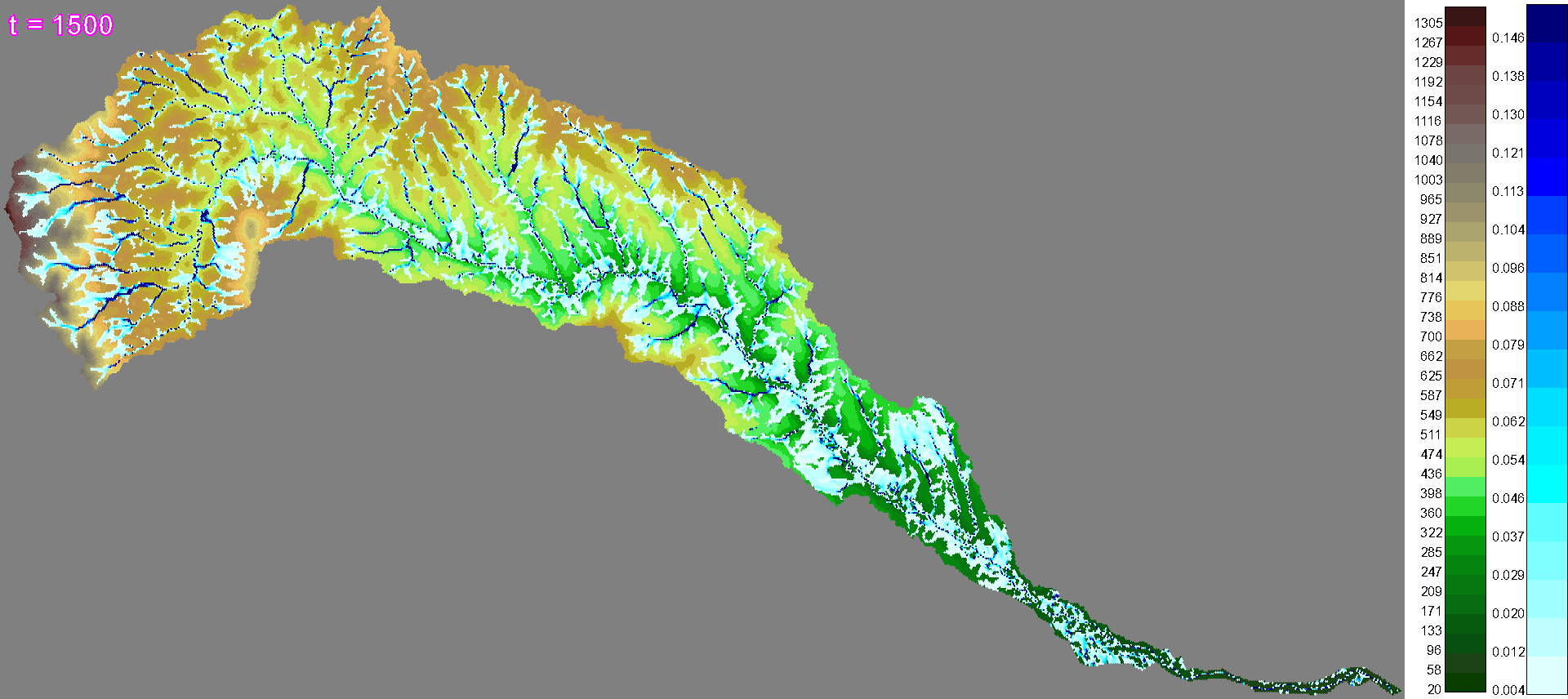}
  \\\vspace{0.5em}
  \includegraphics[width=0.95\textwidth]{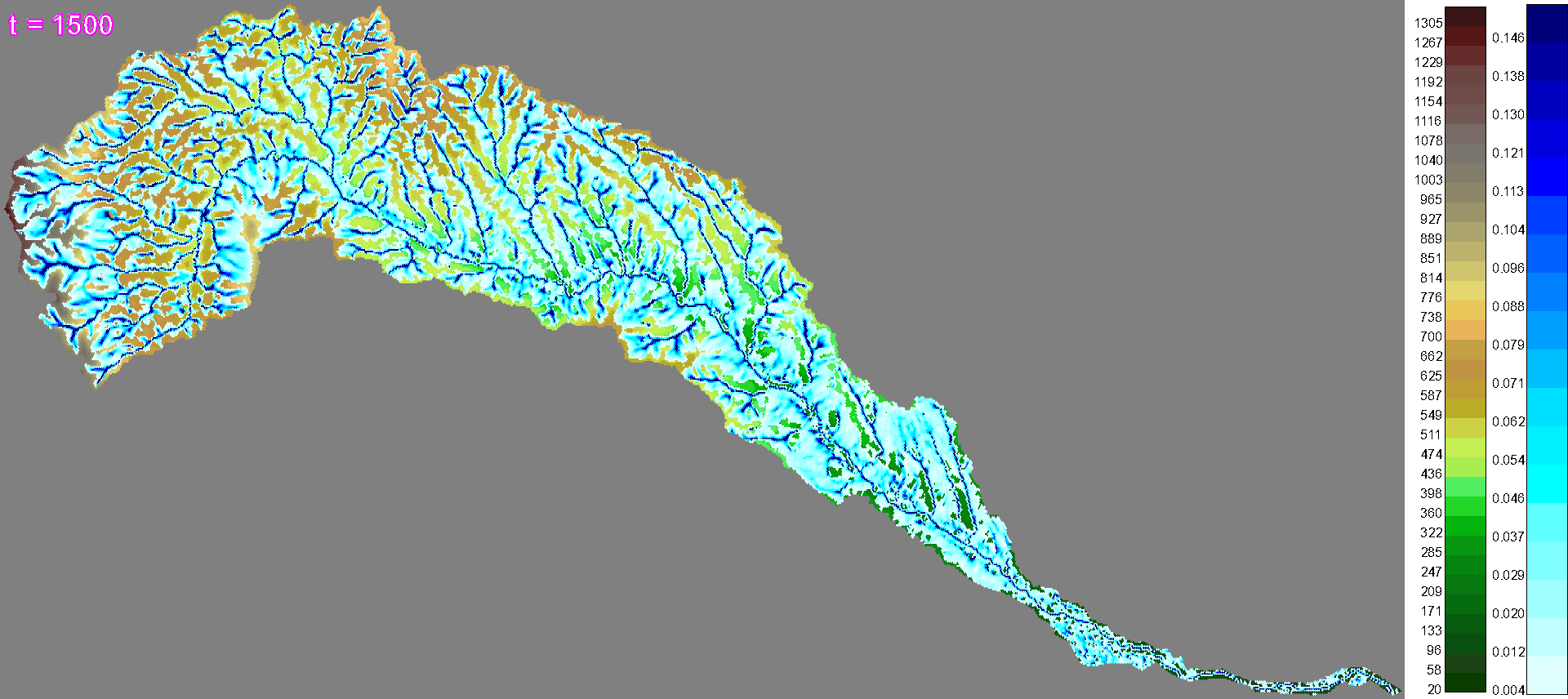}
  \caption{Snapshots of water depth distribution on
    \c{S}u\c{s}i\c{t}a's hydrographic basin at
    $t=1500\; {\rm s}$ for two different uniform vegetation
    densities: $\theta=1.0$ and $\theta=0.97$ on the upper
    and bottom picture, respectively.  The first and the
    second column bar next to the figures are the color
    palettes used for the relief and water depth,
    respectively.  As expected, our numerical data are
    consistent with terrain observations: the amount of
    water leaving the basin is greater in the case of lower
    vegetation density.}
    \label{fig_2Dcomparison_water}
\end{figure}

\begin{figure}[!b]
  \centering
  \includegraphics[width=0.95\textwidth]{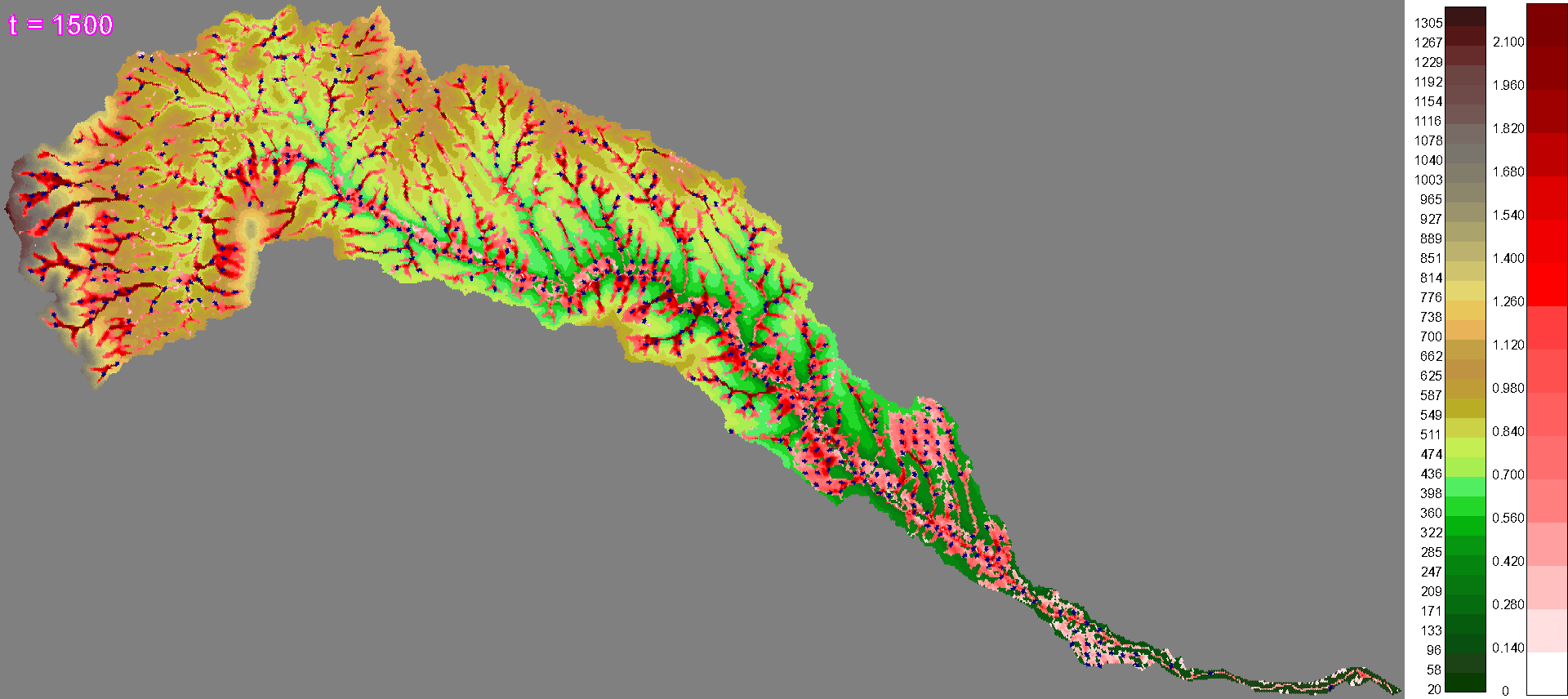}
  \\\vspace{0.5em}
  \includegraphics[width=0.95\textwidth]{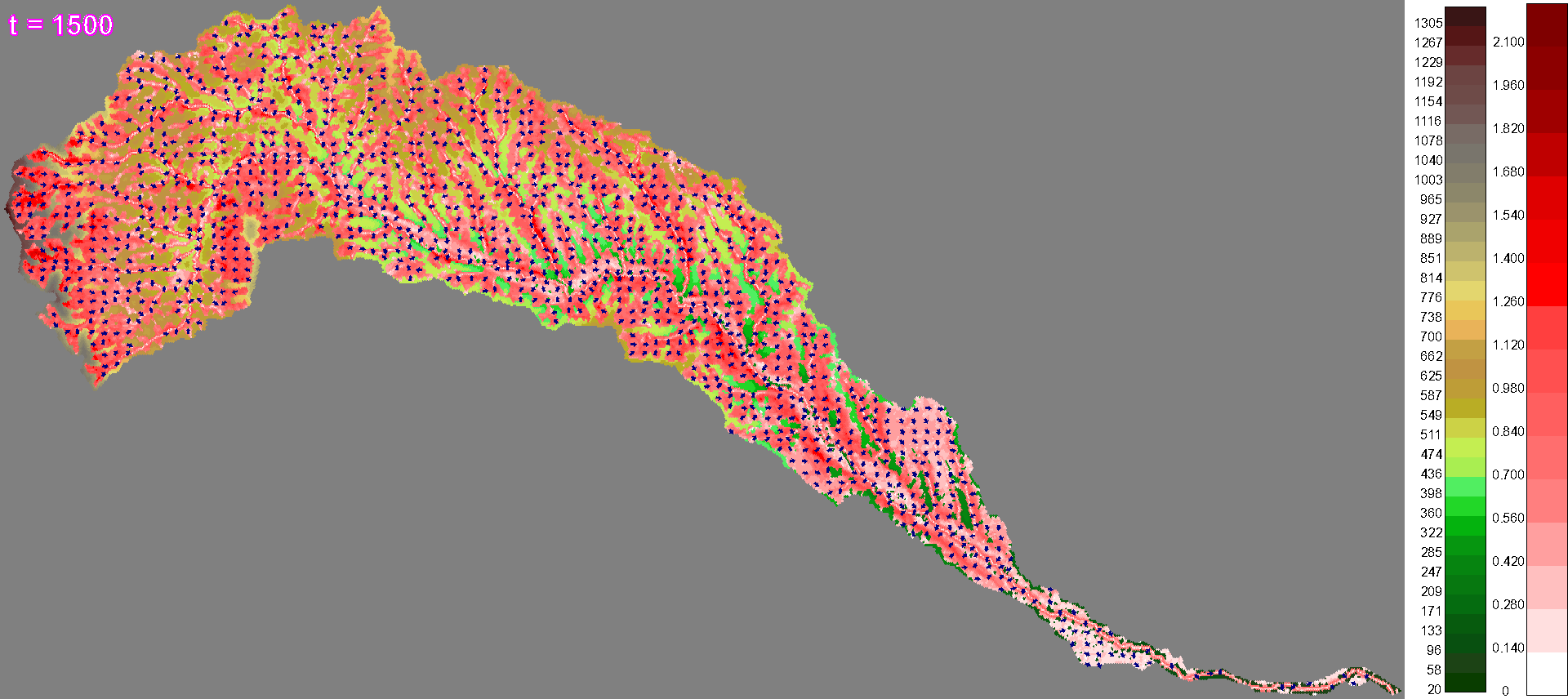}
  \caption{Snapshots of water velocity distribution on
    \c{S}u\c{s}i\c{t}a's hydrographic basin at
    $t=1500\; {\rm s}$ for two different uniform vegetation
    densities: $\theta=1.0$ and $\theta=0.97$ on the upper
    and bottom picture, respectively.  The first and the
    second column bar next to the figures are the color
    palettes used for the relief and modulus of the water
    velocity, respectively.  As expected, the velocities are
    smaller when vegetation is denser.}
    \label{fig_2Dcomparison_velocity}
\end{figure}

The water depth distribution at a given moment in time is
comparatively presented in Fig.~\ref{fig_2Dcomparison_water}
using a blue color gradient (the darkest blue corresponds to
the cells with \mbox{$\theta h \geq 4 h_0$}, where
\mbox{$h_0=i_a*T_d/(2*1000) = 0.0366\,{\rm m}$} is the
rainfall depth).  The water velocity distribution at the
same moment in time is pictured in
Fig.~\ref{fig_2Dcomparison_velocity} using a red color
gradient for the modulus of the velocity and arrows for the
flow direction.  The convention of coloring the relief where
$\theta h \leq 0.1 h_0$ is applied for both figures.  The
data in Table~\ref{table_BasinSusita} confirms that more
water is retained in the basin if the vegetation is denser.
\begin{table}[h]
    \caption{Data for the theoretical experiment on
    \c{S}u\c{s}i\c{t}a's hydrographic basin.  ASTERIX
    was run for $t=3000$ s.}
  \centering
  \begin{tabular}{ cccc }
    \toprule
    $\theta$ & {Rainfall} & {Water out} & {Water left} \\
     -     & {[m$^3$]} & {[m$^3$]} & {[m$^3$]} \\
    \midrule
    $1.00$ & $13166903$ & $104416$ & $13062487$ \\
    $0.97$ & $13166903$ & $90808$  & $13076095$ \\
    \bottomrule
  \end{tabular}
  \label{table_BasinSusita}
\end{table}

\begin{figure}[!htbp]
  \centering
  \begin{tabular}{ ccc }
    \begin{turn}{90}\hspace{48mm}{Rainfall rate [mm/s]}\end{turn}
       &\includegraphics[width=0.55\textwidth]{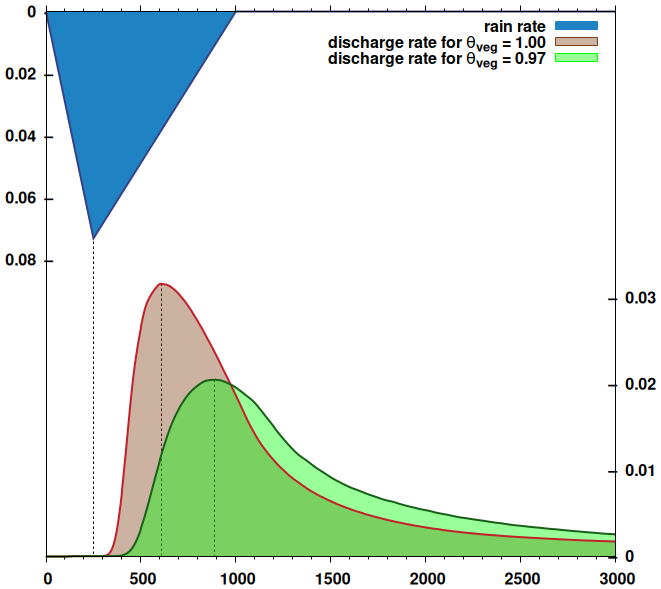}
       &\begin{turn}{90}\hspace{7mm}{Discharge rate [m$^3$/s]}\end{turn}\\
    {} &Time [s] &
  \end{tabular}
  \caption{Simulation of a rainfall and the discharge rates
    on \c{S}u\c{s}i\c{t}a's hydrographic basin for two
    different uniform vegetation densities ($\theta=0.97$
    and $\theta=1$).}
    \label{fig_rain_and_discharge_on_susita}
\end{figure}

\noindent
Furthermore, Fig.~\ref{fig_rain_and_discharge_on_susita}
emphasizes that the maximal discharge rate is smaller and
occurs later when vegetation is denser.

\begin{figure}[!htbp]
  \centering
  \includegraphics[width=0.7\textwidth]{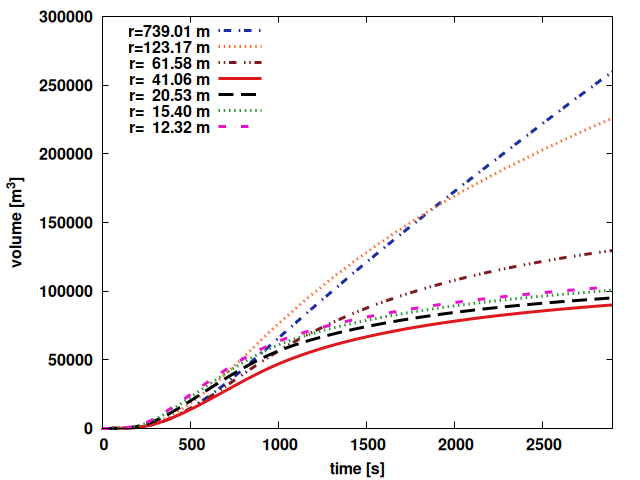}
  \caption{The evolution of the total water leaving the
    basin over time for different grid resolutions.}
  \label{fig_susita_discharge}
\end{figure}

Fig.~\ref{fig_susita_discharge} illustrates the amount of
water leaving the basin over time for different mesh sizes.
One can observe that the curves are getting closer to each
other as the grid resolution increases.  This suggests a
convergence property of the numerical solution to the true
solution of the model.

\begin{remark}
  One can easily test ASTERIX (without having to modify the
  code) for a terrain given by a GIS data file
  (e.g. susita.asc).  Assuming that the source code is
  already compiled, launch from the 'main' directory

  \$ ./natural\_relief\_selection.sh susita

  \$ ./2D\_flow

  \noindent
  The reader is referred to the User Guide for extended
  details.
\end{remark}

\subsubsection{Flow over a slope with vegetation}
\label{sect_flow_over_a_slope_with_veg}
The flow on vegetated areas is more complex than on bare
soil, which may explain the limited availability of reported
data.  In this section, we compare the numerical results
provided by ASTERIX with those reported in
\cite{Dupuis2016}.

The experimental installation consists of a rectangular
laboratory flume ($18$ m long and $1$ m wide) with a
longitudinal bottom slope $S = 1.05$ mm/m and partially
covered with vegetation.  The vegetation is modelled using
uniformly distributed emergent circular cylinders of radius
$R_c=5\, {\rm mm}$ and has a density of
$N = 81\, {\rm cylinders}/{\rm m}^2$.  This type of
vegetation allows the estimation of the porosity by
$\theta_v =1-N\pi R_c^2 \approx 0.99364$.  Six experiments
were performed for different given values of the upstream
flow rate $q_L$, free downstream discharge, two types of
vegetation distribution,
\begin{center}
$
\theta_1= \left\{
  \begin{array}{ll}
    1,& x\in[0,9)\\
    \theta_v,& x\in[9,18]
  \end{array}
\right. ,\quad \quad
\theta_2= \left\{
  \begin{array}{ll}
    \theta_v,& x\in[0,9)\\
    1,& x\in[9,18]
  \end{array}
\right. , 
$
\end{center}
and $z=-m (x-18), \, x\in[0,18]$, with $m=0.105$ mm/m.  The
values of $q_L$ for the first three experiments where
vegetation is present on the lower half of the channel
($\theta=\theta_1$) are
$$q_{11}=7\, {\rm l/s}, \quad q_{21}=15\, {\rm l/s}, \quad q_{31}=21\, {\rm l/s},$$
while the ones for the last three experiments where
vegetation is present on the upper half of the channel
($\theta=\theta_2$) are
$$q_{12}=7\, {\rm l/s}, \quad q_{22}=15\, {\rm l/s}, \quad q_{32}=50\, {\rm l/s}.$$
The numerical simulations are carried out using ASTERIX with
the following boundary conditions
\begin{equation*}
  uh(t,0)=q_L, \quad \partial_x u(t,18)=\partial_x h(t,18)=0,
\end{equation*}
and the following values of the friction parameters
\begin{equation*}
  \alpha_s=0.00709 \quad {\rm and} \quad \alpha_p=73.39.
\end{equation*}

\begin{figure}[!htbp]
  \centering
  \begin{tabular}{ ccc }
    {} & \hspace{5mm}\small{$\theta=\theta_1$} &
         \hspace{2mm}\small{$\theta=\theta_2$}\\
    \begin{turn}{90}\hspace{3mm}\small{Water depth [mm]}\end{turn}\hspace{-5mm}
       & \includegraphics[width=0.24\linewidth]{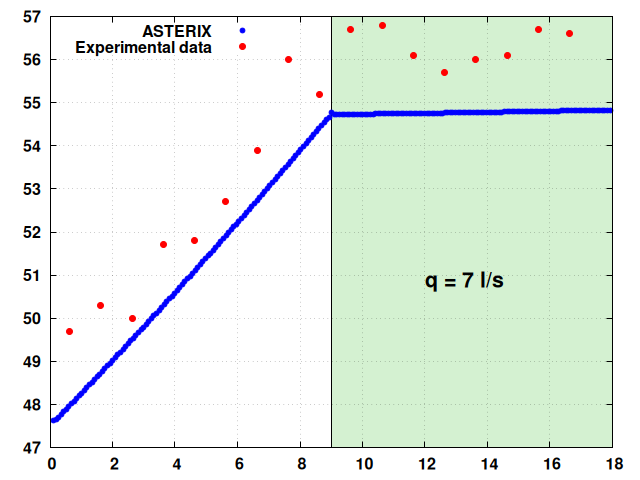}\hspace{-5mm}
       & \includegraphics[width=0.24\linewidth]{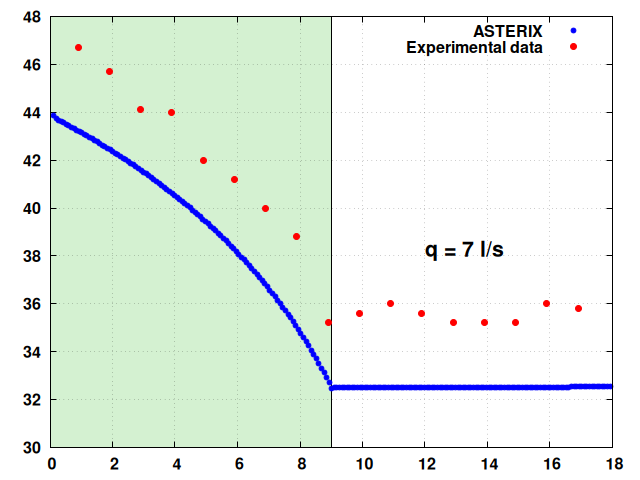}\hspace{-5mm}\\
    \begin{turn}{90}\hspace{3mm}\small{Water depth [mm]}\end{turn}\hspace{-5mm}
       & \includegraphics[width=0.24\linewidth]{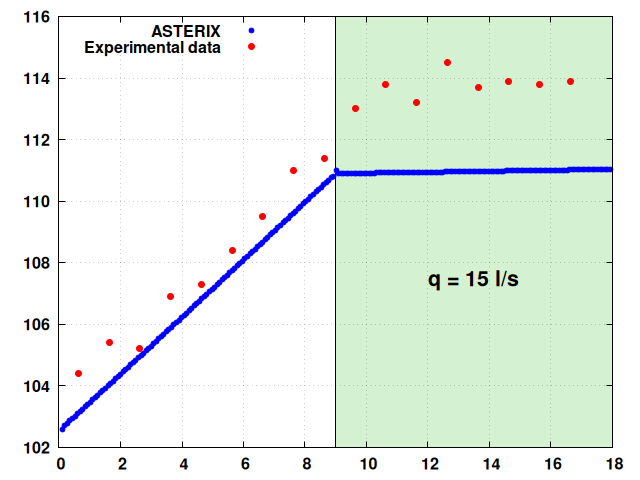}\hspace{-5mm}
       & \includegraphics[width=0.24\linewidth]{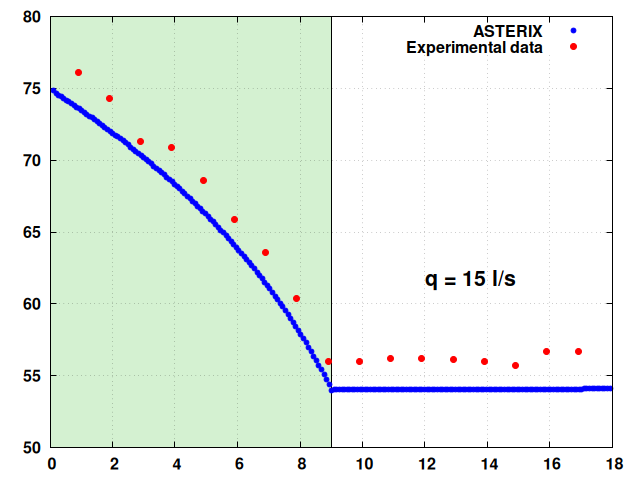}\hspace{-5mm}\\
    \begin{turn}{90}\hspace{3mm}\small{Water depth [mm]}\end{turn}\hspace{-5mm}
       & \includegraphics[width=0.24\linewidth]{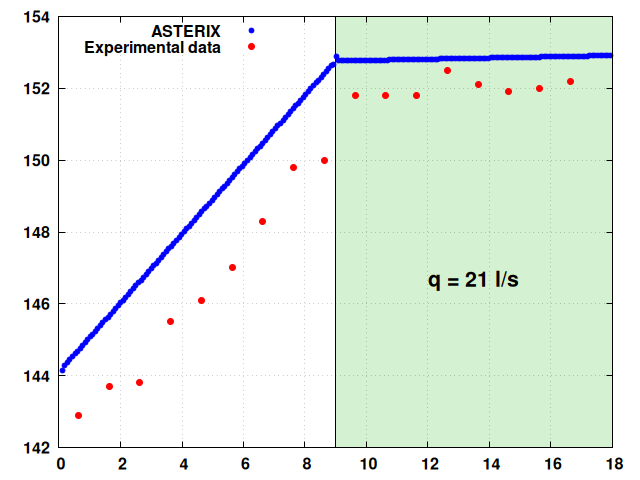}\hspace{-5mm}
       & \includegraphics[width=0.24\linewidth]{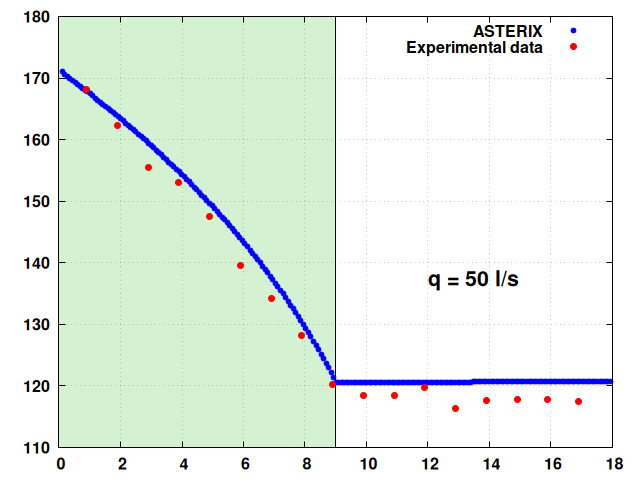}\hspace{-5mm}\\
    {} & \hspace{9mm}\small{x [m]} & \hspace{9mm}\small{x [m]}\\
  \end{tabular}
  \caption{Longitudinal profiles of water depth for the flow over a slope. The vegetated and the bare parts of the soil are marked with green and white, respectively.}
  \label{fig_flow_slope}
\end{figure}

A comparison between the measured data and numerical water
depth is illustrated in Fig.~\ref{fig_flow_slope}.
For each experiment $l$, the error $\epsilon$ (in [mm]) between
the two solutions is calculated with
\begin{equation*}
  \epsilon:=\sqrt{\frac{1}{N_g}\sum_{i=1}^{N_g}\displaystyle\left(h^{\rm exp}(x_i^{l})-h(x_i^{l})\right)^2}, \quad\quad l=\overline{1,6},
  \label{eq_error}
\end{equation*}
where $N_g = 17$ is the number of gauges placed inside the
channel for measurements, $x_i^l$ is the position of the
gauge $G_i$ on the $l^{\rm th}$ experiment, while
$h^{\rm exp}$ and $h$ are the measured and calculated water
depths of the stationary flow, respectively.
Table~\ref{table_errors_6experims} presents the values of
these errors.
\begin{table}[h]
    \caption{The errors between the numerical solution and the experimental data pictured in Fig.~\ref{fig_flow_slope}}
  \centering
  \begin{tabular}{ c|ccc|ccc }
    \toprule
    $\theta$ & \multicolumn{3}{c|}{$\theta_1$} & \multicolumn{3}{c}{$\theta_2$}\\
    \hline
    $q_L$ & $q_{11}$ & $q_{12}$ & $q_{13}$ & $q_{21}$ & $q_{22}$ & $q_{23}$\\
    \midrule
    $\epsilon$ [mm] & 1.48 & 2.05 & 1.73 & 3.08 & 2.10 & 2.73\\
    \bottomrule
  \end{tabular}
  \label{table_errors_6experims}
\end{table}

\begin{figure}[!htbp]
  \centering
  \begin{tabular}{cc}
    \includegraphics[width=0.48\linewidth]{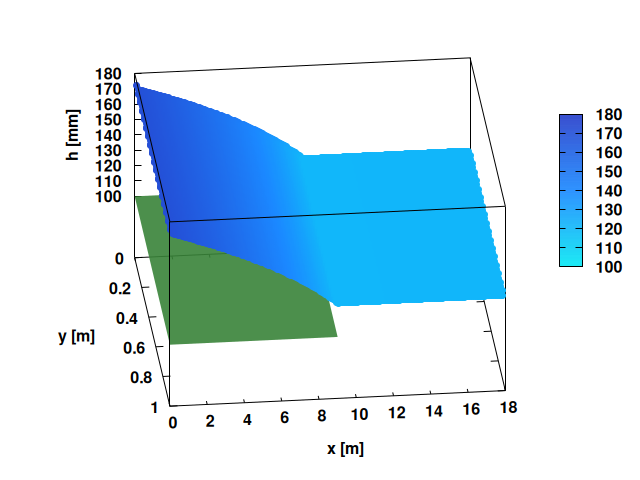}&
    \includegraphics[width=0.48\linewidth]{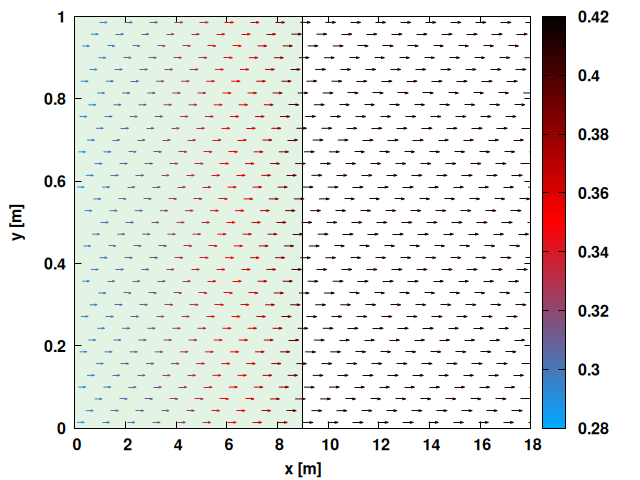}
  \end{tabular}
  \caption{Flow over a slope with vegetation: water depth
    [mm] and velocity field [m/s] obtained with ASTERIX for
    the sixth experiment where $\theta=\theta_2$ and
    $q_L = q_{23}$.}
  \label{fig_flow_slope_veg_2D}
\end{figure}

Fig.~\ref{fig_flow_slope_veg_2D} illustrates the water depth
and the velocity field on the entire channel for the last
experiment from Table~\ref{table_errors_6experims} where
$\theta=\theta_2$ and $q_L = q_{23}$.  The water depth
presented here corresponds to the longitudinal profile found
on the third row and second column in
Fig.~\ref{fig_flow_slope}.
One can observe that the water depth (and velocity)
longitudinal profile is the same for any longitudinal
section along the channel.  Thus, the hexagonal grid used by
the software does not induce numerical adverse effects such
as additional head losses due to the impossibility of the
hexagons to align exactly with the channel walls.

\subsection{Sensitivity analysis}
\label{sect_sensitivity}
One of the main issues in mathematical modelling of real
life phenomena is the sensitivity of the solution of the
model to the variation of the model parameters.  Roughly
speaking, the {\it sensitivity analysis} highlights how
large is the variation of the solution with respect to the
variation of the parameters.  There is a large variety of
methods dedicated to sensitivity analysis; some of them
perform very well (in the sense that they offer relevant
results) for a class of models, but they become inadequate
for other classes.  A very rich literature is devoted to
this subject, see \cite{saltelli_book, saltelli201929}, for
example, for a comprehensive view of the sensitivity analysis
concepts and methods.

Regarding the vegetated Saint-Venant model, we consider the
water-soil $\alpha_s$ (as in the Darcy-Weisbach law) and the
water-plant $\alpha_p$ friction coefficients as the
parameters of the model.  The porosity
$\theta(\boldsymbol{x})$ and the altitude
$z(\boldsymbol{x})$ functions identify a physical
configuration and are considered to be time independent.
The dependence of the hydro-dynamical variables on the
physical configuration is obvious and will not be the
subject of sensitivity analysis; it is rather a problem of
ability of the vegetated Saint-Venant model and numerical
ASTERIX software to offer a reliable solution to a given
hydrological problem.  We note that our analysis is only a
sketch, a systematic analysis being beyond the scope this
article.

\medskip\noindent {\it Sensitivity with respect to the
  discharge $q$, porosity $\theta_v$ and slope $m$}

We proceed and make a theoretical analysis for the response
of the averaged water depth value along the vegetated zone
$9\leq x\leq 18$ to the variation of the environmental
variables $q$, $\theta_v$ and $m$.  Let us introduce
$\overline{h}\left((\alpha_s,\alpha_p);(q,\theta_v,m)\right)$
by
\begin{equation}
  \overline{h}\left((\alpha_s,\alpha_p);(q,\theta_v,m)\right):=\frac{1}{N_v}\sum_i h\left(x_i;(\alpha_s,\alpha_p);(q,\theta_v,m)\right),
  \label{eq_h_bar}
\end{equation}
where $N_v$ is the number of grid points $x_i\geq 9$ taken
inside the vegetated zone,
$h\left(x_i;(\alpha_s,\alpha_p);(q,\theta_v,m)\right)$ is
the steady water depth value given by the numerical model
evaluated at $x_i$ with the following environmental
variables: discharge $q_L=q$, the porosity $\theta_v$ and
the slope $m$.  One defines a regular grid inside the local
domain $\Lambda$:
\begin{equation*}
  \alpha^i_s=\underline{\alpha_s}+(i-1)\triangle_s,\quad\quad
  \alpha^j_p=\underline{\alpha_p}+(j-1)\triangle_p,
\end{equation*}
for $i=\overline{1,N_s}$ and $j=\overline{1,N_p}$.
The sensitivity is quantified by:
\begin{equation}
  \begin{split}
    \delta_s\overline{h}(\boldsymbol{p}) &:=\displaystyle \frac{1}{(N_s-1)N_p} \sum_{i=1}^{N_s-1}\sum_{j=1}^{N_p} \displaystyle\frac{\overline{h}((\alpha^{i+1}_s,\alpha^j_p);\boldsymbol{p})-
      \overline{h}((\alpha^i_s,\alpha^j_p);\boldsymbol{p})}{\triangle_s},\\
    \delta_p\overline{h}(\boldsymbol{p})&:=\displaystyle \frac{1}{N_s(N_p-1)} \sum_{i=1}^{N_s}\sum_{j=1}^{N_p-1} \displaystyle\frac{\overline{h}((\alpha^i_s,\alpha^{j+1}_p);\boldsymbol{p})-\overline{h}((\alpha^i_s,\alpha^{j}_p);\boldsymbol{p})}{\triangle_p},
  \end{split}
\end{equation}
where $\boldsymbol{p}$ stands for the environmental
variables $\boldsymbol{p}=(q,\theta_v,m)$.  The two values
of these functions $\delta_s$ and $\delta_p$ for a given
environmental variable $\boldsymbol{p}$ represent some
average measures of the variation of the water depth
$\overline{h}$ defined in (\ref{eq_h_bar}) with respect to
the water-soil $\alpha_s$ and water-plant $\alpha_p$
friction coefficients, respectively.

\begin{figure}[!htbp]
  \centering
  \begin{tabular}{ ccc }
      {} & {\centering A} & B\\
    {\begin{turn}{90}\hspace{17mm}\begin{turn}{270}$\delta_s$\end{turn}\end{turn}}\hspace{-5mm}
       & \includegraphics[width=0.42\textwidth, height=3.8cm]{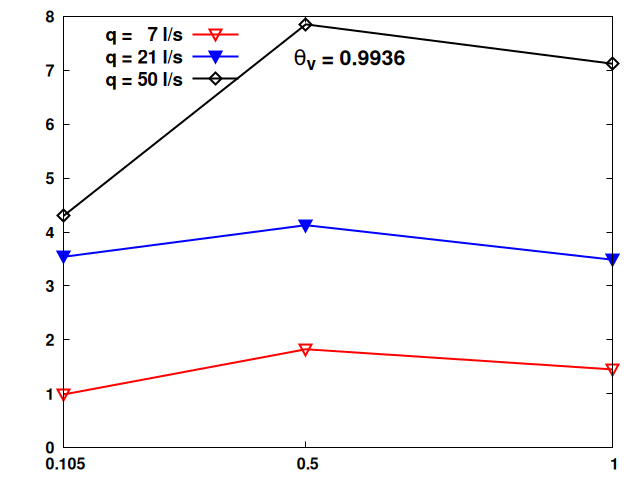}\hspace{-5mm}
       & \includegraphics[width=0.42\textwidth, height=3.8cm]{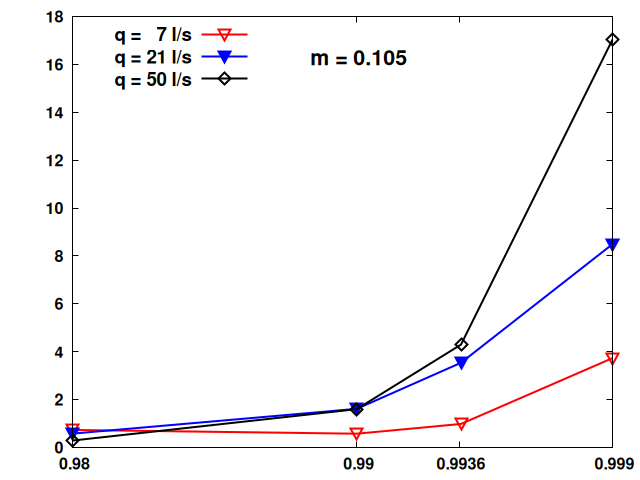}\\
      {} & {\centering C} & D\\
    {\begin{turn}{90}\hspace{19mm}\begin{turn}{270}$\delta_p$\end{turn}\end{turn}}\hspace{-5mm}
       & \includegraphics[width=0.42\textwidth, height=3.8cm]{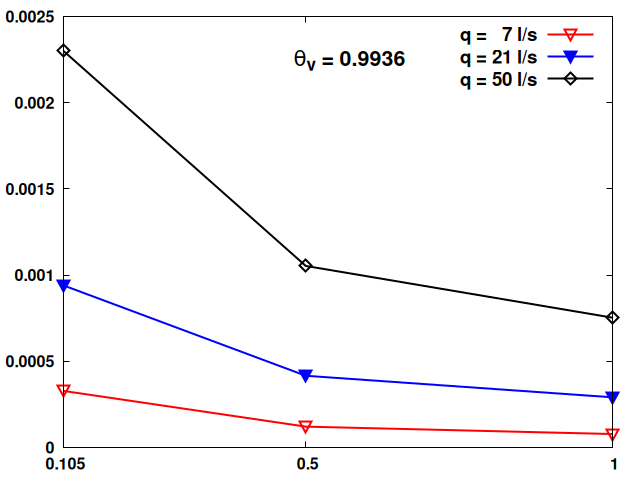}\hspace{-5mm}
       & \includegraphics[width=0.42\textwidth, height=3.8cm]{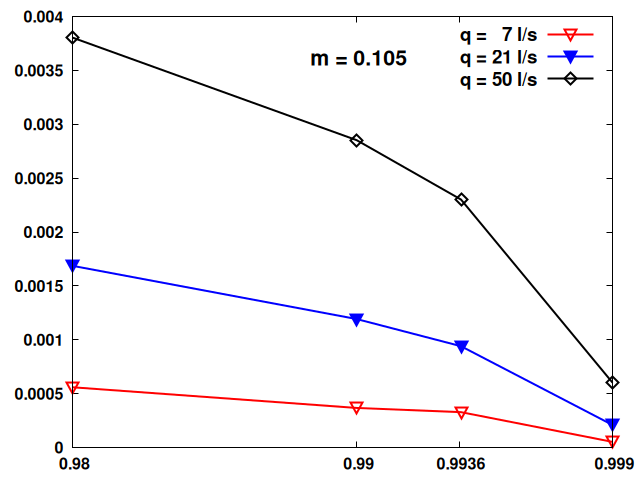}\\
    {} & {$m\,[\%]$} & {$\theta_v$}\\
  \end{tabular}
  \caption{Sensitivities $\delta_s$ (top row) and $\delta_p$
    (bottom row) of the model with respect to the
    environmental variables $\boldsymbol{p}=(q,\theta_v,m)$.
    The values of the sensitivities on the first column are
    calculated with constant porosity $\theta_v$.  The
    values of the sensitivities on the second column are
    calculated with constant slope $m$.}
  \label{fig_sensitivity_as_fcn_of_env_vars}
\end{figure}

Fig.~\ref{fig_sensitivity_as_fcn_of_env_vars} illustrates
the behavior of the sensitivities $\delta_s$ (top row) and
$\delta_p$ (bottom row) of the model with respect to the
environmental variables $\boldsymbol{p}$.  
The porosity
$\theta$ is fixed to $\theta_v=0.9936$ on the first column
and the slope $m$ is fixed to the value $0.105$ on the
second column.  A first observation we can make is that, in
general, the sensitivity increases with the discharge $q$.
An exception can be noticed for $\delta_s$ at small values
of $\theta_v$, see the picture B.  As expected, a lower
vegetation density leads to higher sensitivity with respect
to the water-soil friction coefficient and lower sensitivity
with respect to the water-plant friction coefficient, see
pictures B and D.  Also, as the longitudinal slope $m$
increases (the gravitational forces start to dominate), the
sensitivity with respect to the water-plant friction
coefficient diminishes, see picture C.  We can also observe
that the sensitivity with respect to the water-soil friction
coefficient increases for small values of $m$ and decreases
for higher values of $m$, see picture A.

\medskip\noindent
{\it Estimation of friction coefficients}

Parameter estimation of a physical model is of crucial
importance to ensure the accuracy of the model.  The choice
of an algorithm for solving this problem depends on the
model and the available observational data.  For example, in
\cite{Iber-PEST}, the authors propose an automatic
calibration method to estimate some parameters in a fully
distributed shallow water equation system.  The method is
based on the Gauss-Marquardt-Levenberg algorithm to minimize
a cost function and assumes the evaluation of the first and
second order derivatives of the objective function.  Since
the analytic evaluation is almost impossible, a discrete
derivative evaluation must be developed.  Here is the place
where the sensitivity of the model with respect to
parameters comes into play.

The parameters $\alpha_s$ and $\alpha_p$ can not be directly
measured, they must be estimated by other methods.
Therefore, it is very important to know how its imprecise
assessment affects the solution of the Saint-Venant
equation.  To find these parameters, we use here an inverse
method and the measurements of an experiment reported in
\cite{Dupuis2016}.

To illustrate the difficulties of estimating these friction
parameters, we minimize the errors between the measured and
calculated water depth on a $21\times 26$ grid inside the
domain
\begin{equation*}
  \Lambda:=
  \displaystyle
  \{
  (\alpha_s,\alpha_p)\big|\; 0.0025\leq\alpha_s\leq0.0185,\; 55 \leq\alpha_p\leq 80
  \}.
\end{equation*}
In this sense, we define the error function
\begin{equation*}
  \chi(\alpha_s,\alpha_p):=\sqrt{\frac{1}{N_e N_g}\sum_{l=1}^{N_e}\sum_{i=1}^{N_g}\displaystyle\left(h^{\rm exp}(x_i^{l})-h(x_i^{l};\alpha_s,\alpha_p)\right)^2},
  \label{eq_chi_general}
\end{equation*}
where $N_e$ is the number of considered experiments, $N_g$
is the number of gauges placed inside the channel for
measurements, $x_i^l$ is the position of the gauge $G_i$ on
the $l^{\rm th}$ experiment, while $h^{\rm exp}$ and $h$ are
the measured and calculated water depths, respectively.  For
example, if all six experiments (with $N_g=17$ gauges)
reported in \cite{Dupuis2016} are used, then the error takes
the following form:
  \begin{equation}
    \chi(\alpha_s,\alpha_p) = 
    \sqrt{\frac{1}{102}
      \sum_{j,k,i} \displaystyle\left(h^{\rm exp}(x_i^k;q_{jk};\theta_k)-h(x_i^k;q_{jk};\theta_k,\alpha_s,\alpha_p)\right)^2} .
    \label{eq_chi_6experim}
  \end{equation}
with
$$j=\overline{1,3}, \qquad k=\overline{1,2}, \qquad i=\overline{1,17},$$
and
$$q_{11}=7\, {\rm l/s}, \qquad q_{21}=15\, {\rm l/s}, \qquad q_{31}=21\, {\rm l/s},$$
$$q_{12}=7\, {\rm l/s}, \qquad q_{22}=15\, {\rm l/s}, \qquad q_{32}=50\, {\rm l/s}.$$

For solving a minimization problem, we use the
``Optimization by pattern search'' algorithm
\cite{sherif_and_boice} based on the ``Direct Search''
method introduced in \cite{hooke_and_jeeves}.  We consider
the following cases: \medskip

{\bf Case I}: Experiment with $\theta = \theta_1$, $q=q_{21}$;

{\bf Case II}:  Experiments with $\theta = \theta_1$, $q \in \{q_{11},q_{21},q_{31}\}$;

{\bf Case III}: Experiments with $\theta = \theta_2$, $q \in \{q_{12},q_{22},q_{32}\}$;

{\bf Case IV}: All six experiments: $\theta = \theta_1$,
$q \in \{q_{11},q_{21},q_{31}\}$ and $\theta = \theta_2$,
$q \in \{q_{12},q_{22},q_{32}\}$.
\medskip

\begin{figure}[!htbp]
  \centering
  \includegraphics[width=0.7\textwidth]{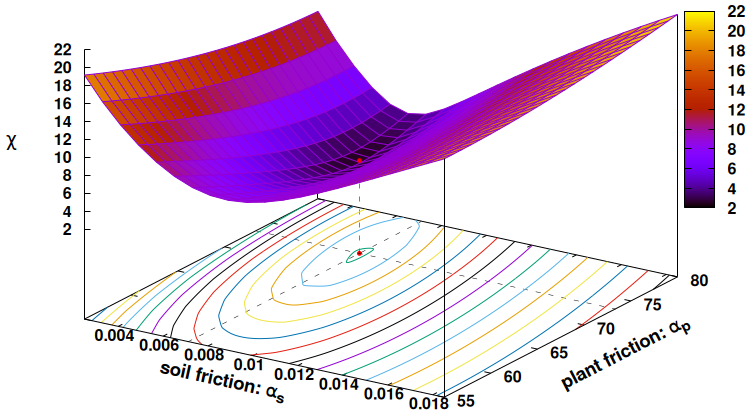}
  \caption{The error surface for Case IV.  The minimum of this function is reached at $\alpha_s=0.00709, \, \alpha_p=73.39$: \\\centering${\rm min}=\chi\left(0.00709,73.39\right)=2.26$ mm.}
  \label{fig_errors_surface}
\end{figure}

When using only one experiment, e.g. {\bf Case I}, the
minimum is poorly located: there are many pairs
$(\alpha _s,\alpha _p)$ in the area of minimum error and it
is not possible to decide which one is the best.  For {\bf
  Case II} and {\bf Case III}, each with 3 experiments, the
parameters are not yet sufficiently well located:
$\alpha _s$ is poorly located in {\bf Case II} and
$\alpha _p$ is poorly located in {\bf Case III}.  For {\bf
  Case IV}, the surface of the error function
(\ref{eq_chi_6experim}) is drawn in
Fig.~\ref{fig_errors_surface} and the minimum is the better
located:
\begin{equation}
  \alpha_s=0.00709 \quad {\rm and} \quad \alpha_p=73.39.
  \label{eq_alfs_alfp_best}
\end{equation}

\section{Conclusions and final remarks}
\label{section_conclusions}
This paper describes a new module of ASTERIX dedicated to
compute the solution of Saint-Venant equations in the
presence of vegetation for water flow modelling.  Based on a
first-order numerical scheme, the computational effort is
reasonable and this makes it suitable to be used from
laboratory level up to practical applications for
environmental problems where the flow surface can be quite
large.

The software is designed to work on regular spatial
hexagonal network.  The software provides a ``porting data''
module \cite{sds-ADataPortingTool} for generating a
hexagonal raster by using data from a rectangular raster or
analytical functions.  The hexagonal raster can also be used
for purposes other than supporting grid for numerical
scheme, e.g flow routing or cellular automata.  The core of
the software consists of two modules: porting data numerical
integration.  Both modules were designed based on two
principles: as simple and as ``universal'' as possible.  By
universal we mean that the software can be used for a large
class of physical environments: topography, land use,
hydrographic basins, soil physics, meteorology, etc.  We
keep it simple in order to allow one to add new modules to
cope with given specific conditions or requirements, e.g.
boundary conditions varying in time and space, distributed
parameters or a post-processing data module.  The source
code and the user manual contain many details to facilitate
the modification of the existent functions or to introduce
new ones.  The user are free to do all he/she wishes.

A lot of work was spent to create a video interface that
allows the visualization of the evolution of the
hydrodynamic variables over time.  This facility can be
informative for hydrologist or environmental policymaker.

Another purpose of the software is to be used as a part of
an inverse method designed to estimate some physical
parameters of the mathematical model.  In this sense, the
numerical integration module has been planned to run
independently from the other modules.  The
Subsection~\ref{sect_sensitivity} presents an example of an
inverse method to estimate the coefficients $\alpha_s$ and
$\alpha_p$ based on a laboratory experiment.  Unfortunately,
the problem of parameter estimation by an inverse method is
ill-posed in the sense that very poor localization of
minimal point of the cost function or amplification of the
error in data are reached.  For example, we found that both
parameters $\alpha_s$ and $\alpha_p$ are very poorly located
when the functional cost is built up on a single experiment,
but they are very well located when one use more and
different experiments (Case IV).

It should be noted that, in addition to the distribution of
the plant cover, the configuration of the soil surface and
the water discharge for which the measurements are made are
important for the determination of $\alpha_s$ and
$\alpha_p$.  This is also highlighted by the theoretical
investigation of the sensitivity of the model with respect
to $q$, $\theta_v$ and $m$ from the second part of
Subsection~\ref{sect_sensitivity}, as can also be noted from
Fig.~\ref{fig_sensitivity_as_fcn_of_env_vars}.

We can also imagine other scenarios (with variable $\theta$
or different values of the slope $m$, for example), but we
did not go further because we wanted to remain inside the
frame of experimental measurements.  This investigation
suggests that one must cleverly choose a set of experiments
with different configurations in order to determine the
parameters as best as possible.

In real conditions, the terrain is heterogeneous and working
only with a single value for $\alpha_s$ and a single value
for $\alpha_p$ can be risky and therefore, distributed
values on the entire domain should be used instead.  In such
cases, a possible solution is to partition the surface into
quasi-homogeneous subdomains with constant friction
parameters.

The estimation of the porosity function is an important
issue from a real application point of view.  For the case
of well structured cover plant (agricultural land, forest or
laboratory experiment), one can easily estimate the area of
domain occupied by the plant stems, but for other cases this
estimation is very difficult.  A possible solution to this
problem is to combine the terrain pointwise determination
with LiDAR measurements or aerial photography.  Another
solution is to set up an inverse method to find the porosity
function.

\medskip\noindent
{\it Future Extensions}

$\circ$ ASTERIX does not yet take into account the water
infiltration and interception, but an extension in this
sense is in progress.  For example, adapting the mass
sources $\mathfrak{M}$ through algebraic type laws can be a
solution to include these processes.

$\circ$ A module for the erosion and sedimentation processes
based on the Hairsine-Rose model \cite{hairrose, kim,
  sander} is being prepared to be attached to our software
in the near future.

$\circ$ Some spatial distributed quantities are only
available as scattered data affected or not by noise.  In
this sense, we are also considering implementing a module
for treating and porting such data on hexagonal raster.

\section*{CRediT authorship contribution statement}
All three authors contributed equally to this work:
Conceptualization, Methodology, Software, Validation, Formal
analysis, Data Curation, Writing - Original Draft.

\section*{Declaration of competing interest}
The authors declare that they have no known competing
financial interests or personal relationships that could
have appeared to influence the work reported in this paper.


\appendix \appendixpage
\section{Details on the Numerical Scheme}
\label{appendix_solve_hv}
The numerical scheme (\ref{fvm_2D_eq_frac.07}) is the result of
combining the following systems from \cite{sds_apnum}
  \begin{subequations}
    \begin{align}
      \sigma_i(\theta_i h_i)^{*}&=\sigma_i(\theta_i h_i)^n
                                  +\triangle t_n{\cal L}_i((h,\boldsymbol{v})^{n}),
                                  \label{fvm_2D_eq_frac.05-1}\\
      \sigma_i(\theta_i h_iv_{a\;i})^{*}&=\sigma_i(\theta_i h_i v_{a\;i})^{n}
                                          +\triangle t_n {\cal J}^v_{a\;i}((h,\boldsymbol{v})^{n})+
                                          \triangle t_n {\cal S}_{a\;i}((h,w)^{n}),
                                          \label{fvm_2D_eq_frac.05-2}
    \end{align}
  \end{subequations}
  \begin{subequations}
    \begin{align}
      \sigma_i(\theta_i h_i)^{n+1}&=\sigma_i(\theta_i h_i)^{*}
                                    +\sigma_i\triangle t_n\mathfrak{M}_i(t^{n+1},h^{n+1}),
                                    \label{fvm_2D_eq_frac.06-1}\\
      \sigma_i(\theta_i h_iv_{a\;i})^{n+1}&=\sigma_i(\theta_i h_iv_{a\;i})^{*}
                                            -\triangle t_n{\cal K}_i(h^{n+1})|\boldsymbol{v}_i^{n+1}| v^{n+1}_{a\;i}.
                                            \label{fvm_2D_eq_frac.06-2}
    \end{align}
  \end{subequations}

\noindent
with $a=1,2$. To get the solution of
(\ref{fvm_2D_eq_frac.07}), one must solve

$\bullet$ the scalar nonlinear equation for $h^{n+1}_i$ obtained by
  combining (\ref{fvm_2D_eq_frac.05-1}) and
  (\ref{fvm_2D_eq_frac.06-1}), and

$\bullet$ the 2D nonlinear system of equations for velocity
  $\boldsymbol{v}^{n+1}_i$ obtained by combining
  (\ref{fvm_2D_eq_frac.05-2}) and
  (\ref{fvm_2D_eq_frac.06-2}).

\noindent
Solving (\ref{fvm_2D_eq_frac.07}) is accomplished in three
steps:
\begin{enumerate}
\item Set $\triangle t_n$ subject to the restriction
   $$\triangle t_n<\tau_n,$$ where
   \begin{equation}
     \tau_n=CFL\displaystyle\frac{\phi_{\rm min}}{c^n_{\rm max}},
     \label{fvm_2D_eq_frac.12}
   \end{equation}
   with
   \begin{equation}
     \begin{array}{c}
       c_i=|\boldsymbol{v}|_i+\sqrt{gh_i}, \quad c_{\rm max}=\max\limits_i \{c_i\},\\
       \phi_{\rm min}=\min\limits_i
       \left\{ \displaystyle\frac{\sigma_i}{\sum\limits_{j\in{\cal N}(i)}l_{(i,j)}} \right\}.
     \end{array}
     \label{fvm_2D_eq_frac.13}
   \end{equation}
 \item Calculate $h_i^{n+1}$ using the mass balance
   relation of (\ref{fvm_2D_eq_frac.07}).
 \item Solve the 2D nonlinear system of equations in
   $\boldsymbol{v}_i^{n}$ given by linear momentum relations
   of (\ref{fvm_2D_eq_frac.07}).
\end{enumerate}

The time step limitation is imposed by the hyperbolic
character of the shallow-water equations and by the
positivity-preserving requirement of the water depth.  $CFL$
is a number between $0$ and $1$ (the Courant-Friedrichs-Lewy
condition).  The reader is referred to
Appendix~\ref{appendix_solve_hv} for extended details on the
last two steps.

When the software is used to study the Riemann Problem or
models that develop shock type discontinuity, it is
recommended \cite{veque, kurganov} to use an augmented
linear momentum flux ${\cal J}^{\boldsymbol{v}}_{a\,i}$
instead of ${\cal J}_{a\,i}(h,\boldsymbol{v})$:
\begin{equation}
  {\cal J}^{\boldsymbol{v}}_{a\,i}:={\cal J}_{a\,i}(h,\boldsymbol{v})+
  \displaystyle\sum\limits_{j\in{\cal N}(i)}l_{(i,j)} \nu_{a(i,j)}(h,\boldsymbol{v}).
  \label{eq_artif_visc}
\end{equation}
This update is based on the artificial viscosity
\begin{equation}
  \nu_{a(i,j)}(h,\boldsymbol{v}) := c_{(i,j)}\mu_{(i,j)} \left( (v_a)_j - (v_a)_i \right),
  \label{eq_visc1}
\end{equation}
where
\begin{equation}
  c_{(i,j)}=\max\{c_i,c_j\}, \qquad 
  \mu_{(i,j)}=\displaystyle \frac{2 \theta_i h_i \theta_j h_j}{\theta_i h_i + \theta_j h_j}.
\end{equation}

This last step has an analytic solution of the form
\begin{equation}
\boldsymbol{v}^{n+1}_i 
= \mu\cdot (\theta_i h_i\boldsymbol{v}_i)^{*},
\end{equation}
where
\begin{equation}
  \mu = \frac{2}{\alpha +\sqrt{\alpha
      ^2+4\beta|\boldsymbol{\gamma |}}},
  \label{miu-formula}
\end{equation}
with
$$
\alpha := (\theta_i h_i)^{n+1},\qquad
\beta := \triangle t_n{\cal K}(h_i^{n+1},\theta_i), \qquad
\gamma _a := (\theta_i h_iv_{a\;i})^{*}.
$$

Note that for the particular cases when the mass source
$\mathfrak{M}$ does not depend on $h$ (e.g. only the rain is
taken into consideration), the computation of $h^{n+1}_i$ is
direct and does no longer require to solve a nonlinear
equation.

\section{Analytic Solution for Thacker's Problem}
\label{sect_apendix_thacker_nonoscilatoriu}
We present here a particular analytic solution of the
shallow-water equations (\ref{swe_vegm_rm.02}) in the
absence of vegetation ($\theta = 1$), for
$\mathfrak{t}^{p} = \boldsymbol{0}$ and
$\mathfrak{t}^{s} = h \tau \boldsymbol{v}$, where $\tau$ is
a proportionality coefficient, for a soil surface
$z(\boldsymbol{x})$ of the form
\begin{equation}
  z(x,y) = a (x-x_0)^2 + b (y-y_0)^2, \quad a,b>0,
\end{equation}
and for a water velocity which does not depend on space
variable,
i.e. $\boldsymbol{v}(t,\boldsymbol{x})=\boldsymbol{f}(t)$.
To avoid confusions in formulas, we consider
\begin{equation}
  \boldsymbol{x}=(x,y)^T \quad {\rm and} \quad \boldsymbol{v}=(u,v).
\end{equation}
A damped non-oscillating solution is obtained if
\begin{equation}
  \Delta_a := \tau ^2 - 8ga >0 \quad {\rm and} \quad \Delta_b := \tau ^2 - 8gb >0.
  \label{eq_cond_osc}
\end{equation}
In order to write the general solution of the flow, we
introduce the notations:
\begin{equation}
  u_0 := u(0), \quad u'_0 := u'(0), \quad v_0 := v(0), \quad v'_0 := v'(0),
\end{equation}
\begin{equation}
  \lambda _1 = \lambda _1^{\Delta} := \frac{-\tau - \sqrt{\Delta }}{2}, \quad 
  \lambda _2 = \lambda _2^{\Delta} := \frac{-\tau + \sqrt{\Delta }}{2},
\end{equation}
\begin{equation}
  A = A^{\Delta} := \frac{V_0\lambda _2-V'_0}{\sqrt{\Delta }},\quad 
  B = B^{\Delta} := -\frac{V_0\lambda _1-V'_0}{\sqrt{\Delta }},
\end{equation}
\begin{equation}
  V(t; \Delta , V_0, V'_0) := Ae^{\lambda _1t}+Be^{\lambda _2t},
\end{equation}

\begin{equation}
  W_0(t; \Delta , V_0, V'_0) := \frac{1}{2g}
  \left [
    A^2\frac{\lambda _2}{\lambda _1}
    \left (
      1-e^{2\lambda _1t}
    \right )+
    B^2\frac{\lambda _1}{\lambda _2}
    \left (
      1-e^{2\lambda _2t}
    \right )+
    2AB
    \left (
      1-e^{-\tau t}
    \right )
  \right ],
\end{equation}

\begin{equation}
    w_0(t):= w_0(t;\Delta _a,\Delta _b,u_0,v_0,u'_0,v'_0)
    := w_0+ W_0(t; \Delta _a, u_0, u'_0)+ W_0(t; \Delta _b, v _0, v'_0).
\end{equation}

The general solution can now be written as
\begin{equation}
  \left \{
    \begin{aligned}
      h(t,x,y) & =[w(t,x,y)-z(x,y)]_+,\\
      u(t) & = V(t;\Delta _a, u_0, u'_0),\\
      v(t) & = V(t;\Delta _b, v_0, v'_0),
    \end{aligned}
  \right .
  \label{eq_thacker_sol_2d}
\end{equation}
where $[\alpha]_+$ is the positive part of $\alpha$ and
\begin{equation}
  w(t,x,y) := w_0(t) -\frac{1}{g}(x-x_0)(\tau u(t) + u'(t))-\frac{1}{g}(y-y_0)(\tau v(t) + v'(t))
\end{equation}
is the evolution of the free water surface.

\noindent {\bf Note: } A damped oscillating solution is
obtained if both inequalities (\ref{eq_cond_osc}) are not
satisfied; if only one of the inequalities is not satisfied,
then the solution is oscillating only in one direction.

\section{Rainfall Hyetograph: triangular model}
\label{sect_apendix_rainfall_triangular}
\begin{figure}[!htbp]
  \centering
  \includegraphics[width=0.40\linewidth]{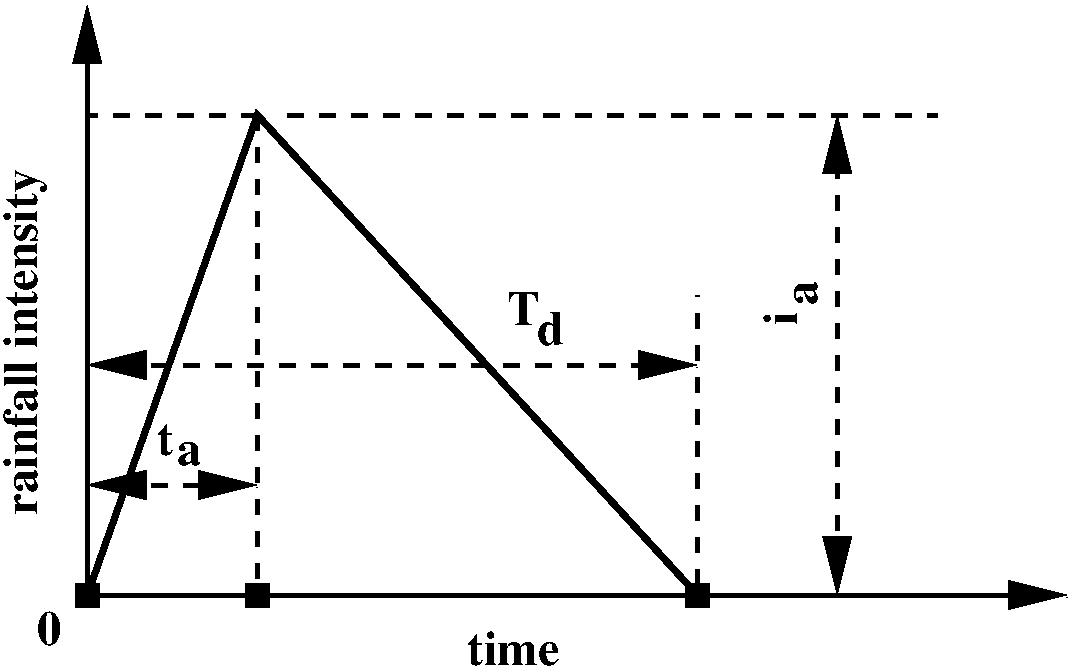}
  \caption{Hyetograph of a rainfall: triangular model.}
  \label{fig_hyetograph_triangle}
\end{figure}

This rainfall model has three parameters: the total time
$T_d$ of the rain, the maximum intensity $i_a$ of the rain
and the time $t_a$ at which this maximum intensity is
recorded.  The ratio between the total time $T_d$ and the
moment of time $t_a$ is called the rainfall advance
coefficient.
$$ 
\gamma=\frac{t_a}{T_d}.
$$
The graph of the instantaneous rain intensity $r(t)$ 
$$
r(t)=\left\{
  \begin{array}{ll}
    i_a\displaystyle\frac{t}{t_a},& 0 \leq t\leq t_a,\\
    i_a\displaystyle\frac{1-t/T_d}{1-\gamma},& t_a<t \leq T_d,\\
    0,& {\rm otherwise}
  \end{array}
\right.
$$
over the time interval $[0,T_d]$ has the shape of a triangle 
with $T_d$ and $i_a$ as its base and height, respectively (see
Fig.~\ref{fig_hyetograph_triangle}).

The amount of water falling over the interval
$[t_1,t_2]$ is:
$$
V(t_1,t_2)=\int\limits_{t_1}^{t_2}r(t){\rm d}t,
$$
and therefore, the total amount of water falling over the
interval $[0,T_d]$ is equal to the area of the triangle:
$$
V=\frac{i_a T_d}{2}.
$$

\bibliographystyle{elsarticle-num}
\bibliography{biblio_ammwfvh}

\begin{thebibliography}{10}
\expandafter\ifx\csname url\endcsname\relax
  \def\url#1{\texttt{#1}}\fi
\expandafter\ifx\csname urlprefix\endcsname\relax\def\urlprefix{URL }\fi
\expandafter\ifx\csname href\endcsname\relax
  \def\href#1#2{#2} \def\path#1{#1}\fi

\bibitem{worldbank_water_res_manag}
{World Bank},
  \href{https://www.worldbank.org/en/topic/waterresourcesmanagement}{Water
  resources management} (2022).
\newline\urlprefix\url{https://www.worldbank.org/en/topic/waterresourcesmanagement}

\bibitem{unep_disasters_and_climate_change}
{United Nations Environment Programme},
  \href{https://www.unep.org/topics/fresh-water/disasters-and-climate-change}{Disasters
  and climate change} (2024).
\newline\urlprefix\url{https://www.unep.org/topics/fresh-water/disasters-and-climate-change}

\bibitem{oecd_book}
OECD,
  \href{https://www.oecd-ilibrary.org/content/publication/ed1a7936-en}{Toolkit
  for Water Policies and Governance}, OECD Publishing, Paris, 2021.
\newblock \href {http://dx.doi.org/10.1787/ed1a7936-en}
  {\path{doi:10.1787/ed1a7936-en}}.
\newline\urlprefix\url{https://www.oecd-ilibrary.org/content/publication/ed1a7936-en}

\bibitem{land_cover_01}
D.-H. Kim, J.~M. Johnson, K.~C. Clarke, H.~K. McMillan, Untangling the impacts
  of land cover representation and resampling in distributed hydrological model
  predictions, Environmental Modelling \& Software 172 (2024) 105893.
\newblock \href {http://dx.doi.org/10.1016/j.envsoft.2023.105893}
  {\path{doi:10.1016/j.envsoft.2023.105893}}.

\bibitem{land_cover_02}
Y.~Zhou, Y.~Zhang, J.~Vaze, P.~Lane, S.~Xu, Improving runoff estimates using
  remote sensing vegetation data for bushfire impacted catchments, Agricultural
  and Forest Meteorology 182-183 (2013) 332--341.
\newblock \href {http://dx.doi.org/10.1016/j.agrformet.2013.04.018}
  {\path{doi:10.1016/j.agrformet.2013.04.018}}.

\bibitem{orth}
R.~Orth, M.~Staudinger, S.~I. Seneviratne, J.~Seibert, M.~Zappa, Does model
  performance improve with complexity? {A} case study with three hydrological
  models, Journal of Hydrology 523 (2015) 147--159.
\newblock \href {http://dx.doi.org/10.1016/j.jhydrol.2015.01.044}
  {\path{doi:10.1016/j.jhydrol.2015.01.044}}.

\bibitem{schoups2008}
G.~Schoups, N.~C. van~de Giesen, H.~H.~G. Savenije, Model complexity control
  for hydrologic prediction, Water Resources Research 44~(12) (2008) W00B03,
  1--14.
\newblock \href {http://dx.doi.org/10.1029/2008WR006836}
  {\path{doi:10.1029/2008WR006836}}.

\bibitem{beven2006}
K.~Beven, A manifesto for the equifinality thesis, Journal of Hydrology 320~(1)
  (2006) 18--36, the model parameter estimation experiment.
\newblock \href {http://dx.doi.org/10.1016/j.jhydrol.2005.07.007}
  {\path{doi:10.1016/j.jhydrol.2005.07.007}}.

\bibitem{wheater_book}
H.~Wheater, S.~Sorooshian, K.~D. Sharma, Hydrological Modelling in Arid and
  Semi-Arid Areas, Cambridge University Press, 2007.
\newblock \href {http://dx.doi.org/10.1017/CBO9780511535734}
  {\path{doi:10.1017/CBO9780511535734}}.

\bibitem{epa_rainfall_runoff}
J.~Sitterson, C.~Knightes, R.~Parmar, K.~Wolfe, M.~Muche, B.~Avant, An overview
  of rainfall-runoff model types, Tech. rep., U.S. Environmental Protection
  Agency, USA (2017).

\bibitem{R_source}
I.~P. Sam~Albers, \href{https://CRAN.R-project.org/view=Hydrology}{Cran task
  view: Hydrological data and modeling} (2024).
\newline\urlprefix\url{https://CRAN.R-project.org/view=Hydrology}

\bibitem{Py_source}
R.~Collenteur,
  \href{https://github.com/raoulcollenteur/Python-Hydrology-Tools}{Python
  hydrology tools} (2024).
\newline\urlprefix\url{https://github.com/raoulcollenteur/Python-Hydrology-Tools}

\bibitem{hype}
C.~Brendel, R.~Capell, A.~Bartosova, Rational gaze: Presenting the open-source
  {HYPE}tools {R} package for analysis, visualization, and interpretation of
  hydrological models and datasets, Environmental Modelling \& Software 178
  (2024) 106094.
\newblock \href {http://dx.doi.org/10.1016/j.envsoft.2024.106094}
  {\path{doi:10.1016/j.envsoft.2024.106094}}.

\bibitem{abbott_SHE}
M.~Abbott, J.~Bathurst, J.~Cunge, P.~O'Connell, J.~Rasmussen, An introduction
  to the european hydrological system — systeme hydrologique europeen,
  ``she'', 1: History and philosophy of a physically-based, distributed
  modelling system, Journal of Hydrology 87~(1) (1986) 45--59.
\newblock \href {http://dx.doi.org/10.1016/0022-1694(86)90114-9}
  {\path{doi:10.1016/0022-1694(86)90114-9}}.

\bibitem{mike_she}
{DHI},
  \href{https://www.dhigroup.com/technologies/mikepoweredbydhi/mike-she}{Mike
  she} (2024).
\newline\urlprefix\url{https://www.dhigroup.com/technologies/mikepoweredbydhi/mike-she}

\bibitem{woolhiser_kineros}
D.~Woolhiser, R.~Smith, D.~Goodrich, U.~A.~R. Service, KINEROS: A Kinematic
  Runoff and Erosion Model: Documentation and User Manual, 1 disc with report,
  U.S. Department of Agriculture, Agricultural Research Service, ARS-77, 1989.

\bibitem{liang_VIC}
X.~Liang, D.~P. Lettenmaier, E.~F. Wood, S.~J. Burges, A simple hydrologically
  based model of land surface water and energy fluxes for general circulation
  models, Journal of Geophysical Research: Atmospheres 99~(D7) (1994)
  14415--14428.
\newblock \href {http://dx.doi.org/10.1029/94JD00483}
  {\path{doi:10.1029/94JD00483}}.

\bibitem{singh_book}
V.~P. Singh, \href{https://www.wrpllc.com/books/cmwhn.html}{Computer Models of
  Watershed Hydrology}, Water Resources Publications, LCC, 2012.
\newline\urlprefix\url{https://www.wrpllc.com/books/cmwhn.html}

\bibitem{delestre}
O.~Delestre, C.~Lucas, P.-A. Ksinant, F.~Darboux, C.~Laguerre, et~al.,
  {SWASHES}: a compilation of shallow water analyticsolutions for hydraulic and
  environmental studies, International Journal for Numerical Methods in Fluids
  72~(3) (2013) 269--300.
\newblock \href {http://dx.doi.org/10.1002/fld.3741}
  {\path{doi:10.1002/fld.3741}}.

\bibitem{sds_apnum}
S.~Ion, D.~Marinescu, S.~G. Cruceanu, Numerical scheme for solving a porous
  {Saint}-{Venant} type model for water flow on vegetated hillslopes, Appl.
  Numer. Math. 172 (2022) 67--98.
\newblock \href {http://dx.doi.org/10.1016/j.apnum.2021.09.019}
  {\path{doi:10.1016/j.apnum.2021.09.019}}.

\bibitem{cadam}
J.~Hiver, {A}dverse-{S}lope and {S}lope ({B}ump), in: S.~Soares-Frazão,
  M.~Morris, Y.~Zech (Eds.), {C}oncerted {A}ction on {D}am {B}reak {M}odelling
  - {CADAM}: Objectives, Project Report, Test Cases, Meeting Proceedings,
  Universit\'{e} Catholique de Louvain, Civ. Eng. Dept., Hydraulic Division,
  Louvain-la-Neuve, Belgium (CD-ROM), 2000, {R}eport {SR} 571 {HR}
  {W}allingford.

\bibitem{Dupuis2016}
V.~Dupuis, S.~Proust, C.~Berni, A.~Paquier, Combined effects of bed friction
  and emergent cylinder drag in open channel flow, Environmental Fluid
  Mechanics 16~(6) (2016) 1173--1193.
\newblock \href {http://dx.doi.org/10.1007/s10652-016-9471-2}
  {\path{doi:10.1007/s10652-016-9471-2}}.

\bibitem{sds-ADataPortingTool}
S.~Ion, D.~Marinescu, S.~G. Cruceanu, V.~Iordache, A data porting tool for
  coupling models with different discretization needs, Environmental Modelling
  \& Software 62 (2014) 240--252.
\newblock \href {http://dx.doi.org/10.1016/j.envsoft.2014.09.012}
  {\path{doi:10.1016/j.envsoft.2014.09.012}}.

\bibitem{rouse}
H.~Rouse, Elementary Mechanics of Fluids Hardcover, John Wiley and Sons, Inc.,
  New York, USA, 1946.

\bibitem{baptist}
M.~J. Baptist, V.~Babovic, J.~R. Uthurburu, M.~Keijzer, R.~E. Uittenbogaard,
  A.~Mynett, A.~Verwey, On inducing equations for vegetation resistance,
  Journal of Hydraulic Research 45~(4) (2007) 435--450.
\newblock \href {http://dx.doi.org/10.1080/00221686.2007.9521778}
  {\path{doi:10.1080/00221686.2007.9521778}}.

\bibitem{nepf}
H.~M. Nepf, Drag, turbulence, and diffusion in flow through emergent
  vegetation, Water Resources Research 35~(2) (1999) 479--489.
\newblock \href {http://dx.doi.org/10.1029/1998WR900069}
  {\path{doi:10.1029/1998WR900069}}.

\bibitem{Whitaker}
S.~Whitaker, Diffusion and dispersion in porous media, AIChE Journal 13~(3)
  (1967) 420--427.
\newblock \href {http://dx.doi.org/10.1002/aic.690130308}
  {\path{doi:10.1002/aic.690130308}}.

\bibitem{Guo2021}
K.~Guo, M.~Guan, D.~Yu, Urban surface water flood modelling - a comprehensive
  review of current models and future challenges, Hydrology and Earth System
  Sciences 25~(5) (2021) 2843--2860.
\newblock \href {http://dx.doi.org/10.5194/hess-25-2843-2021}
  {\path{doi:10.5194/hess-25-2843-2021}}.

\bibitem{Sanders2008}
B.~F. Sanders, J.~E. Schubert, H.~A. Gallegos, Integral formulation of
  shallow-water equations with anisotropic porosity for urban flood modeling,
  Journal of Hydrology 362~(1-2) (2008) 19--38.
\newblock \href {http://dx.doi.org/10.1016/j.jhydrol.2008.08.009}
  {\path{doi:10.1016/j.jhydrol.2008.08.009}}.

\bibitem{Guinot2017DualIP}
V.~Guinot, B.~F. Sanders, J.~E. Schubert, Dual integral porosity shallow water
  model for urban flood modelling, Advances in Water Resources 103 (2017)
  16--31.

\bibitem{SoaresFrazo2018}
S.~Soares-Frazão, F.~Franzini, J.~Linkens, J.-C. Snaps, Investigation of
  distributed-porosity fields for urban flood modelling using single-porosity
  models, E3S Web of Conferences 40 (2018) 06040.
\newblock \href {http://dx.doi.org/10.1051/e3sconf/20184006040}
  {\path{doi:10.1051/e3sconf/20184006040}}.

\bibitem{Varra2024}
G.~Varra, L.~Cozzolino, R.~Della~Morte, S.~Soares-Frazão, Shallow water
  equations with binary porosity and their application to urban flooding,
  Physics of Fluids 36~(7).
\newblock \href {http://dx.doi.org/10.1063/5.0214441}
  {\path{doi:10.1063/5.0214441}}.

\bibitem{strang}
G.~Strang, On the construction and comparison of difference schemes, {SIAM}
  Journal on Numerical Analysis 5~(3) (1968) 506--517.
\newblock \href {http://dx.doi.org/10.1137/0705041}
  {\path{doi:10.1137/0705041}}.

\bibitem{veque-phd}
R.~J. LeVeque, Time-split methods for partial differential equations, Ph.D.
  thesis, Stanford University, Stanford, California, USA (1982).

\bibitem{veque}
R.~J. LeVeque, Finite Volume Methods for Hyperbolic Problems, Cambridge Texts
  in Applied Mathematics, Cambridge University Press, Cambridge, UK, 2002.
\newblock \href {http://dx.doi.org/10.1017/CBO9780511791253}
  {\path{doi:10.1017/CBO9780511791253}}.

\bibitem{sds_atee2023}
S.~Ion, D.~Marinescu, S.~Cruceanu, Analysis of the effects of curvature on the
  solutions of shallow water equations, in: 2023 13th International Symposium
  on Advanced Topics in Electrical Engineering (ATEE), 2023, pp. 1--6.
\newblock \href {http://dx.doi.org/10.1109/ATEE58038.2023.10108117}
  {\path{doi:10.1109/ATEE58038.2023.10108117}}.

\bibitem{thacker}
W.~Thacker, Some exact solutions to the nonlinear shallow-water equations,
  Journal of Fluid Mechanics 107 (1981) 499--508.
\newblock \href {http://dx.doi.org/10.1017/S0022112081001882}
  {\path{doi:10.1017/S0022112081001882}}.

\bibitem{sampson}
J.~J. Sampson, A.~Easton, M.~Singh, Moving boundary shallow water flow in a
  region with quadratic bathymetry, ANZIAM Journal 49 (2007) C666--C680.
\newblock \href {http://dx.doi.org/10.21914/anziamj.v49i0.306}
  {\path{doi:10.21914/anziamj.v49i0.306}}.

\bibitem{matskevich}
N.~Matskevich, L.~Chubarov, Exact solutions to shallow water equations for a
  water oscillation problem in an idealized basin and their use in verifying
  some numerical algorithms, Numerical Analysis and Applications 12 (2019)
  234--250.
\newblock \href {http://dx.doi.org/10.1134/S1995423919030030}
  {\path{doi:10.1134/S1995423919030030}}.

\bibitem{smc-rp-19}
S.~Ion, D.~Marinescu, S.-G. Cruceanu, Constructive approach of the solution of
  the {Riemann} problem for shallow water equations with topography and
  vegetation, An. St. Univ. Ovidius 28~(2) (2020) 93--114.
\newblock \href {http://dx.doi.org/10.2478/auom-2020-0021}
  {\path{doi:10.2478/auom-2020-0021}}.

\bibitem{google_earth_susita}
{Google Earth},
  \href{https://earth.google.com/web/@45.94352578,26.99150326,231.35063202a,61462.69767313d,35y,0h,0t,0r/data=OgMKATA}{Map
  of \c{S}u\c{s}i\c{t}a river basin} (2024).
\newline\urlprefix\url{https://earth.google.com/web/@45.94352578,26.99150326,231.35063202a,61462.69767313d,35y,0h,0t,0r/data=OgMKATA}

\bibitem{saltelli_book}
A.~Saltelli, K.~Chan, E.~Scott, Sensitivity Analysis, Wiley, 2009.

\bibitem{saltelli201929}
A.~Saltelli, K.~Aleksankina, W.~Becker, P.~Fennell, F.~Ferretti, N.~Holst,
  S.~Li, Q.~Wu, Why so many published sensitivity analyses are false: A
  systematic review of sensitivity analysis practices, Environmental Modelling
  \& Software 114 (2019) 29--39.
\newblock \href {http://dx.doi.org/10.1016/j.envsoft.2019.01.012}
  {\path{doi:10.1016/j.envsoft.2019.01.012}}.

\bibitem{Iber-PEST}
G.~Garc{\'i}a-Al{\'e}n, C.~Montalvo, L.~Cea, J.~Puertas, Iber-{PEST}:
  {A}utomatic calibration in fully distributed hydrological models based on the
  2{D} shallow water equations, Environmental Modelling \& Software 177 (2024)
  106047.
\newblock \href {http://dx.doi.org/10.1016/j.envsoft.2024.106047}
  {\path{doi:10.1016/j.envsoft.2024.106047}}.

\bibitem{sherif_and_boice}
Y.~S. Sherif, B.~A. Boice, Optimization by pattern search, European Journal of
  Operational Research 78~(3) (1994) 277--303.
\newblock \href {http://dx.doi.org/10.1016/0377-2217(94)90041-8}
  {\path{doi:10.1016/0377-2217(94)90041-8}}.

\bibitem{hooke_and_jeeves}
R.~Hooke, T.~A. Jeeves, ``direct search'' solution of numerical and statistical
  problems, Journal of the ACM 8~(2) (1961) 212--229.
\newblock \href {http://dx.doi.org/10.1145/321062.321069}
  {\path{doi:10.1145/321062.321069}}.

\bibitem{hairrose}
P.~Hairsine, C.~Rose, Modeling water erosion due to overland flow using
  physical principles: 1. sheet flow, Water Resources Research 28~(1) (1992)
  237--243.
\newblock \href {http://dx.doi.org/10.1029/91WR02380}
  {\path{doi:10.1029/91WR02380}}.

\bibitem{kim}
J.~Kim, V.~Y. Ivanov, N.~D. Katopodes, Modeling erosion and sedimentation
  coupled with hydrological and overland flow processes at the watershed scale,
  Water Resources Research 49~(9) (2013) 5134--5154.
\newblock \href {http://dx.doi.org/10.1002/wrcr.20373}
  {\path{doi:10.1002/wrcr.20373}}.

\bibitem{sander}
G.~Sander, J.-Y. Parlange, D.~Barry, M.~Parlange, W.~L. Hogarth, Limitation of
  the transport capacity approach in sediment transport modeling, Water
  Resources Research 43~(2), w02403.
\newblock \href {http://dx.doi.org/10.1029/2006WR005177}
  {\path{doi:10.1029/2006WR005177}}.

\bibitem{kurganov}
A.~Kurganov, G.~Petrova, A second-order well-balanced positivity preserving
  central-upwind scheme for the {Saint}-{Venant} system, Communications in
  Mathematical Sciences 5~(1) (2007) 133--160.
\newblock \href {http://dx.doi.org/10.4310/CMS.2007.v5.n1.a6}
  {\path{doi:10.4310/CMS.2007.v5.n1.a6}}.

\end{thebibliography}

\end{document}